\documentclass[aps,prx,reprint,amsfonts,amssymb,amsmath,longbibliography]{revtex4-2}

\usepackage{amssymb,amsmath,bm}
\usepackage{fullpage}
\usepackage{graphicx}
\usepackage{wrapfig}
\usepackage{floatflt}
\usepackage{color}
\usepackage{xspace}
\usepackage{dsfont}
\usepackage[official,right]{eurosym}
\usepackage[top=0.8in, bottom=0.8in, left=0.75in, right=0.75in]{geometry}
\usepackage[utf8]{inputenc}
\usepackage{hyperref}

\newcommand{\SUL} {H$_8$C$_4$SO$_2\cdot$Cu$_2$Cl$_4$ \xspace}

\newcommand{\TCL} {TlCuCl$_3$\xspace}

\newcommand{\RbCu}{Rb$_2$Cu$_2$Mo$_3$O$_{12}$\xspace}

\newcommand{\be}{\begin{equation} }
\newcommand{\ee}{\end{equation} }
\newcommand{\bea}{\begin{eqnarray} }
\newcommand{\eea}{\end{eqnarray} }

\begin{document}

\title{Critical dielectric susceptibility at a magnetic BEC quantum critical point}

\author{S.~Hayashida}
\email{shoheih@phys.ethz.ch}
\affiliation{Laboratory for Solid State Physics, ETH Z{\"u}rich, 8093 Z{\"u}rich, Switzerland}
\author{L.~Huberich}
\affiliation{Laboratory for Solid State Physics, ETH Z{\"u}rich, 8093 Z{\"u}rich, Switzerland}
\author{D.~Flavi{\'a}n}
\affiliation{Laboratory for Solid State Physics, ETH Z{\"u}rich, 8093 Z{\"u}rich, Switzerland}
\author{Z. Yan}
\affiliation{Laboratory for Solid State Physics, ETH Z{\"u}rich, 8093 Z{\"u}rich, Switzerland}
\author{K. Yu. Povarov}
\affiliation{Laboratory for Solid State Physics, ETH Z{\"u}rich, 8093 Z{\"u}rich, Switzerland}
\author{S. Gvasaliya}
\affiliation{Laboratory for Solid State Physics, ETH Z{\"u}rich, 8093 Z{\"u}rich, Switzerland}
\author{A. Zheludev}
\email{zhelud@ethz.ch}
\homepage{http://www.neutron.ethz.ch/}
\affiliation{Laboratory for Solid State Physics, ETH Z{\"u}rich, 8093 Z{\"u}rich, Switzerland}

\date{\today}

\begin{abstract}
Magnetic-field-induced phase transitions are investigated in the frustrated gapped quantum paramagnet Rb$_{2}$Cu$_{2}$Mo$_3$O$_{12}$ through dielectric and calorimetric measurements on single-crystal samples.
It is clarified that the previously reported dielectric anomaly at 8~K in powder samples is not due to  a chiral spin liquid state as has been suggested, but rather to a tiny amount of a ferroelectric impurity phase. Two field-induced quantum phase transitions between paraelectric and paramagnetic and ferroelectric and magnetically ordered states are clearly observed.
It is shown that the electric polarization is a secondary  order parameter at the lower-field (gap closure) quantum critical point but a primary one at the saturation transition.
Having clearly identified the magnetic Bose-Einstein condensation (BEC) nature of the latter, we use the dielectric channel to directly measure the critical divergence of BEC susceptibility.
The observed power-law behavior is in very good agreement with theoretical expectations for three-dimensional BEC.
Finally, dielectric data reveal magnetic presaturation phases in this compound that may feature exotic order with unconventional broken symmetries.  
\end{abstract}

\maketitle

\section{Introduction}
Being the basis of superfluidity, superconductivity~\cite{Leggett2006quantum} and numerous other phenomena in systems ranging from cold atoms~\cite{Anderson1995,Davis1995} to semiconductors~\cite{Keldysh1968,Kasprzak2006} to ferromagnetic films~\cite{Demokritov2006}, Bose-Einstein condensation (BEC) is arguably the most celebrated of all phase transitions.
BEC is also a key example of a {\em quantum} phase transition (QPT) that can occur at zero temperature and is driven by quantum fluctuations~\cite{Proukakis2017BEC}.
For lattice gases it represents a transition from a gapless superfluid to a gapped Mott insulator state~\cite{Fisher1989,Greiner2002}.
One key property of any continuous phase transition is a divergent susceptibility.
Unfortunately, in BEC's original formulation the order parameter, namely the complex amplitude of the condensate wave function, has no physical field associated with it.
Hence the critical susceptibility is not even physical, let alone experimentally accessible.
A possible workaround can be found in magnetic insulators.
Field-induced saturation transitions in conventional Heisenberg antiferromagnets~\cite{Batyev1984}, as well as soft-mode ordering transitions in gapped quantum paramagnets~\cite{Giamarchi1999,Nikuni2000,Giamarchi2008,Zapf2014}, can be described in terms of a BEC of magnons.
Here, the BEC order parameter is the spontaneous spatially modulated (often simply staggered) magnetization transverse to the applied field~\cite{Batyev1984,Giamarchi2008}.
The critical susceptibility acquires a concrete physical meaning.

Unfortunately, it still remains inaccessible, as there is no practical way to produce a measurement field that is modulated at the atomic length scale.
At best, it can in principle be inferred from correlation functions measured in scattering experiments~\cite{Lovesey1984}.
In this paper, we demonstrate how the critical susceptibility at a BEC QPT can be measured {\em directly}, by actually applying an excitation field and measuring the response.
We study a quantum magnetic material with a field-induced magnon BEC QPT, in which magnetoelectric coupling makes the {\em uniform  electric polarization} a primary BEC order parameter.
The critical behavior of susceptibility at the quantum critical trajectory can thus be directly studied via dielectric measurements.

The fact that magnetic-field-induced transitions in spin systems may sometimes show dielectric anomalies is well known~\cite{Tokura2014}.
Pioneering measurements of critical susceptibility at a magnetic quantum critical point (QCP) via the dielectric channel were performed at the saturation transition in Ba$_2$CoGe$_2$O$_7$~\cite{Kim2014}.
Here, electric polarization is indeed a primary order parameter, being coupled {\em linearly} to the staggered magnetization via the so-called spin-dependent $p$-$d$ hybridization mechanism~\cite{Arima2007,Jia2007}.
However, the behavior observed in Ba$_2$CoGe$_2$O$_7$ is representative of the Ising, rather than the BEC, universality class~\cite{Kim2014}.

Critical dielectric susceptibility was also observed in magnetic BEC transitions of gapped quantum paramagnets near their {\em lower} critical fields, particularly in the spin ladder compound \SUL~\cite{Schrettle2013,Povarov2015} and the coupled spin-dimer system TlCuCl$_3$~\cite{Kimura2016,Kimura2017}.
Here, the magnetoelectric coupling is believed to be of the ``reverse Dzyaloshinskii-Moriya''~\cite{Mostovoy2006} or ``spin current''~\cite{Katsura2005} origin.
As shown in Ref.~\cite{Povarov2015} and explained below, for these transitions, electric polarization is only critical at the thermodynamic (finite-temperature) phase transition but becomes a {\em secondary} order parameter at $T\rightarrow 0$.
Consequently, dielectric susceptibility is not critical or divergent at the QCP, but merely shows a finite jump.
In this paper, we focus on a different spin-gap material, namely, \RbCu.
This compound has two experimentally accessible magnetic-field-induced magnon BEC QPTs, one at the gap-closing field $H_{c1}$ and another one at saturation field $H_{c2}$.
We show that in the $T\rightarrow 0$ limit, polarization is a primary BEC order parameter at the {\em upper} critical field but not at the lower one.

The peculiar magnetic and dielectric properties of the quasi-one-dimensional frustrated ferro-antiferromagnet \RbCu have been studied for over a decade~\cite{Hase2004,Hase2005,Yasui2014,Hayashida2019,Yasui2013_1,Yasui2013_2,Reynolds2019,Ueda2020}.
The ground state is a gapped quantum paramagnet, with magnetic long-range order induced in a magnon BEC transition at $\mu_{0}H_{c1}\sim 2$~T.
The BEC ``dome'' extends up to $\mu_{0}H_{c2}\sim 13$~T, where the system becomes saturated.
Magnetic long-range order is in all fields confined to $T<1.5$~K.
The material generated a great deal of excitement when it was found to exhibit {\em magnetic}-field-induced ferroelectricity at much higher temperatures, already at $T'=8$~K~\cite{Yasui2013_1,Yasui2013_2,Reynolds2019,Ueda2020}.
It was suggested that below $T'$ the system is a chiral spin liquid, with no magnetic long-range order but with spontaneously chiral spin fluctuations~\cite{Furukawa2010}.
At much lower temperatures, another dielectric anomaly is detected at the boundary of the BEC dome~\cite{Reynolds2019}.

The problem with all those studies is that they were carried out on powder samples.
This limits control over sample quality and is particularly hurtful for any measurements in an applied field, where any features in the data will be smeared out by the directional averaging of magnetic anisotropy.  
In contrast, this paper is based on {\em single crystals}.
We first show that the $T'=8$~K anomaly is {\em spurious} and due to an impurity phase.
We then do what is {\em in principle} impossible in powders: measure the quantum scaling behavior of critical dielectric susceptibility at the $H_{c2}$ magnon BEC QCP.

\section{Material and methods}
\subsection{\RbCu: a brief introduction}
\RbCu crystallizes in a monoclinic structure (space group $C2/c$)~\cite{Solodovnikov1997}.
The magnetic properties are due to $S=1/2$ Cu$^{2+}$ cations.
CuO$_{4}$ plaquettes form a one-dimensional chain along the crystallographic $b$ axis.
Powder samples have been extensively investigated by means of magnetic and dielectric measurements~\cite{Hase2004,Hase2005,Yasui2014,Hayashida2019,Yasui2013_1,Yasui2013_2,Reynolds2019,Ueda2020}, high-pressure studies~\cite{Kuroe2006,Hamasaki2007}, nuclear magnetic resonance (NMR)~\cite{Yagi2017,Matsui2017}, neutron scattering~\cite{Tomiyasu2009,Reynolds2019,Ueda2020}, muon spin relaxation measurements~\cite{Kawamura2018,Reynolds2019}, and electron spin resonance~\cite{Hayashida2019,Ueda2020}.
The ground state is a nonmagnetic singlet with a spin gap $\Delta \sim 2$~K, and the spins become fully polarized at $\mu_{0}H\sim 13$~T~\cite{Yasui2014,Hayashida2019}.
It is believed that the exchange interactions are highly frustrated: The ferromagnetic nearest-neighbor $J_{1}=-138$~K and antiferromagnetic next-nearest-neighbor $J_{2}=51$~K~\cite{Hase2004,Hase2005}.
A recent neutron scattering study suggested an interplay of more complex interactions such as interchain and anisotropic couplings~\cite{Ueda2020}.
Comprehensive magnetothermodynamic measurements were carried out on \RbCu single crystals~\cite{Hayashida2019}.
The entire field-temperature phase diagram was mapped out.
It turned out to be curiously anisotropic: the lower critical fields differing by up to 50{\%} depending on the field direction, and the upper ones being almost the same in all geometries.

\begin{figure}[tbp]
\includegraphics[scale=1]{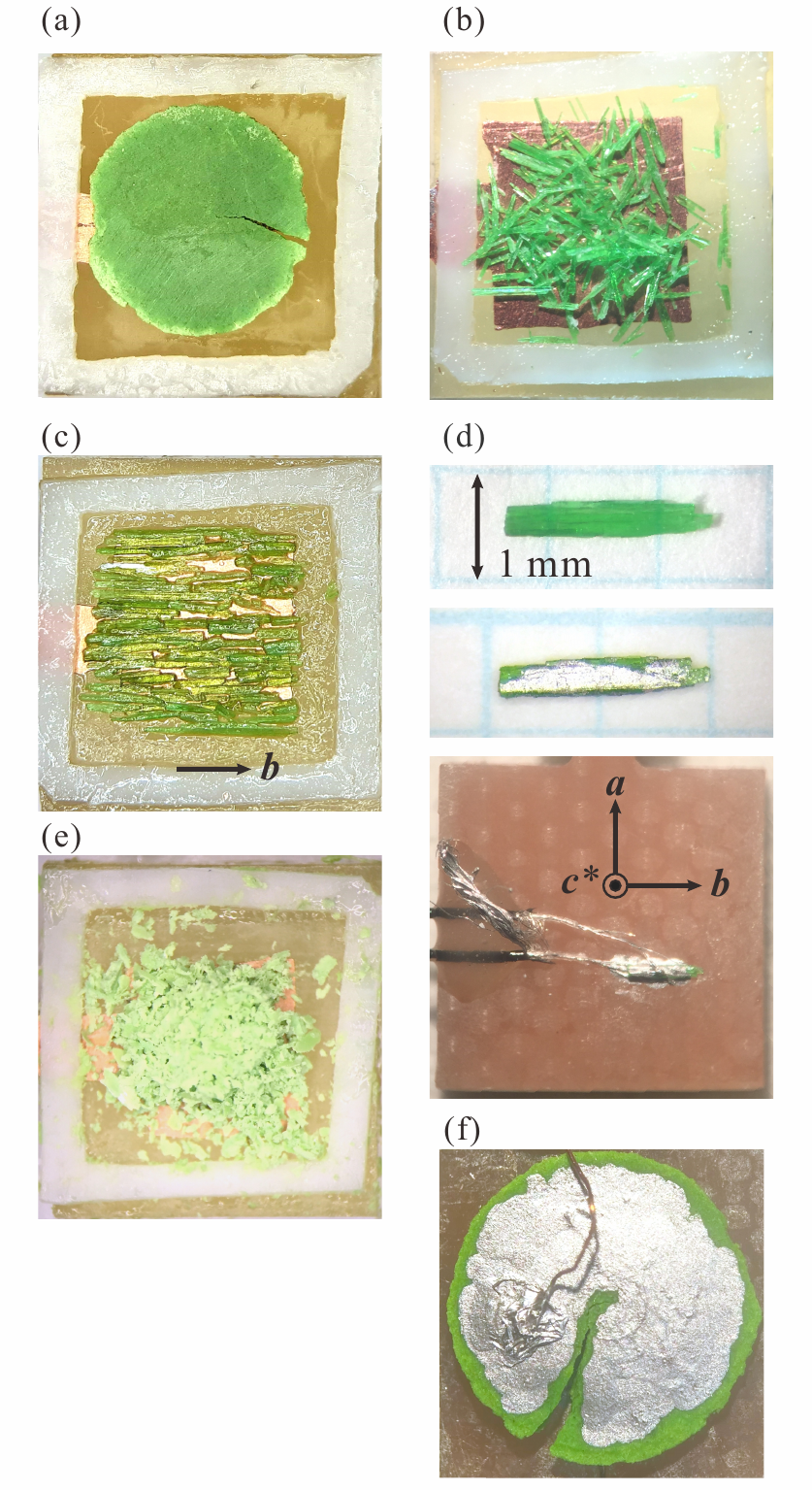}
\caption{Photos of various \RbCu samples used: (a) sintered-powder sample, (b) randomly aligned crystals, (c) coaligned crystals, (d) single-crystal sample without and with silver-paste contacts, (e) finely ground crystals, and (f) pelletized sintered-powder sample with silver-paste contacts attached.}
\label{fig:samples}
\end{figure}

\subsection{Sample synthesis and characterization}
In this paper we employ six types of samples (Fig.~\ref{fig:samples}):
(1) sintered powder prepared following the procedure described in Ref.~\cite{Hase2004} and mounted in a capacitance cell;
(2) an assembly of about 200 randomly aligned small crystals in a capacitance cell; (3) about 50 hand-picked larger single crystals aligned to have their $b$ axes parallel in a capacitance cell; (4) a single crystal of $\sim 0.1$ ~mg with silver-paste contacts;
(5) a powder produced by finely grinding small single crystals, in a capacitance cell; and (6) pelletized sintered-powder sample with silver-paste contacts.
The single crystals were grown by a spontaneous crystallization using a flux method~\cite{Solodovnikov1997}.
They are pale green, translucent, typically 2~mm long, and needle shaped, with the long edge parallel to the crystallographic $b$ direction and well-formed $(10\bar{1})$ or $(001)$ side faces.
Powders and powdered crystals were characterized using powder x-ray diffraction on a Rigaku MiniFlex diffractometer.
Within experimental accuracy, all samples were found to be a single phase and fully consistent with the reported crystal structure~\cite{Solodovnikov1997}.
Single crystals were characterized and aligned using single-crystal x-ray diffraction on a Bruker APEX-II diffractometer.
The structure was again found to be in excellent agreement with that previously published~\cite{Hayashida2019}.

\subsection{Experimental procedures}
For comparative dielectric measurements, samples of types (1)--(3) and (5) were placed in a capacitance cell with plate size $6\times6$~mm$^2$ and spacing of 0.2~mm.
Loaded measurement cells are shown in Figs.~\ref{fig:samples}(a)--\ref{fig:samples}(c) and \ref{fig:samples}(e).
For single-crystal measurements, we used silver-paste contacts directly deposited on opposing $(001)$ surfaces of a $0.3\times1.8\times0.2$~mm$^3$ single crystal, as displayed in Fig.~\ref{fig:samples}(d).
For pyroelectric current measurements, we deposited silver-paste contacts on the surface of a sintered-powder sample that was pelletized by baking at 440~$^{\circ}$C for 60~h. The pellet area was 27~mm$^{2}$, and the thickness was 0.78~mm [Fig.~\ref{fig:samples}(f)].

The capacitance was measured using an Andeen-Hagerling capacitance bridge at a frequency of 1~kHz.
We measured both real, $C'$, and imaginary, $C''$, parts of the capacitance.
An excitation voltage between 1.5 and 15~V was applied.
We used 15~V to map out the phase diagram with as high as possible signal-to-noise ratio and used 3~V to measure critical susceptibility in a linear response regime as discussed below.
Frequency variation of the capacitance was measured by a Keysight E4980A Precision {\it LCR} meter. A direct current (dc) bias voltage of up to 50~V was supplied by a Keithley 6517A electrometer.
The pyroelectric current $I_{p}$ was measured using the same device.

All measurements were carried out in a Quantum Design Physical Property Measurement System (PPMS) with a maximum magnetic field of 14~T.
Low-temperature data down to $T=0.1$~K were taken with a $^3$He-$^4$He dilution refrigerator PPMS insert.
For the capacitance measurements, the temperature and magnetic field were swept continuously with minimum sweeping rates of 0.01~K/min and 2.5~Oe/s, correspondingly.
For measurements of electrical polarization, a voltage of $\pm 250$~V, corresponding to an electric field of $0.32$~kV/mm, was applied when cooling the sample from 1.1 to 0.1~K at 11~T to avoid ferroelectric domain formation. After turning off the voltage at 0.1~K, we waited for about 10 min until any extrinsic current had disappeared. $I_{p}$ was measured during the warming of the sample with a sweeping rate of 0.5~K/min. Electric polarization $P$ is derived from the time integral of the pyroelectric current. The poling electric field was applied parallel to the magnetic field.
Additional heat capacity data for $\mathbf{H}\|\mathbf{a}$ and $\mathbf{c}^{*}$ were collected on a 0.12-mg single-crystal sample using a standard Quantum Design relaxation calorimetry option as measured in a previous study~\cite{Hayashida2019}.

\section{Results and discussion}
\subsection{High-temperature dielectric properties}\label{sec:highT}
\begin{figure}[tbp]
\includegraphics[scale=1]{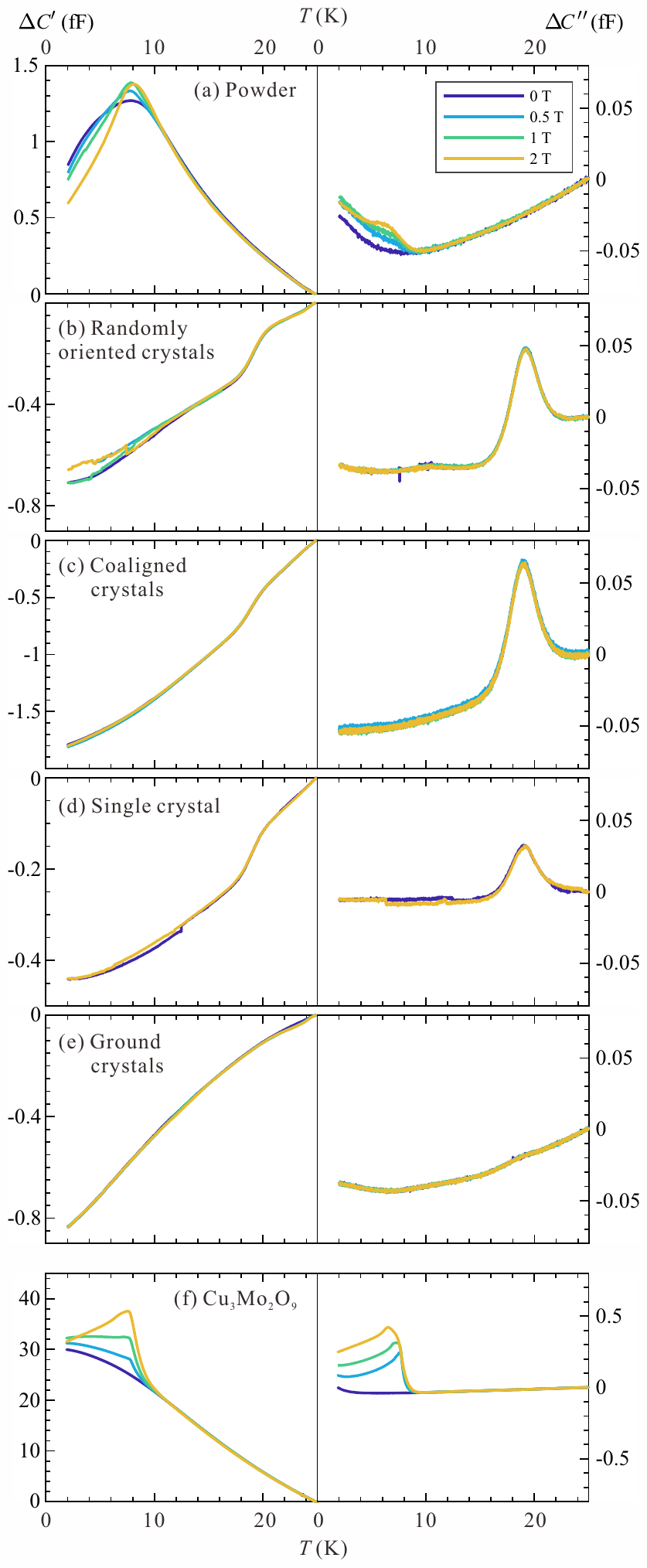}
\caption{Temperature scans of the capacitance above 2~K for (a) sintered powder, (b) randomly oriented crystals, (c) coaligned crystals, (d) a single piece of the crystal, (e) powder produced from finely ground crystals, and (f) Cu$_{3}$Mo$_{2}$O$_{9}$ powder.
The labels of the left and right axes are the changes in the capacitance from 25-K $\Delta C'$ and $\Delta C''$, respectively.}
\label{fig:highT}
\end{figure}
\begin{figure}[tbp]
\includegraphics[scale=1]{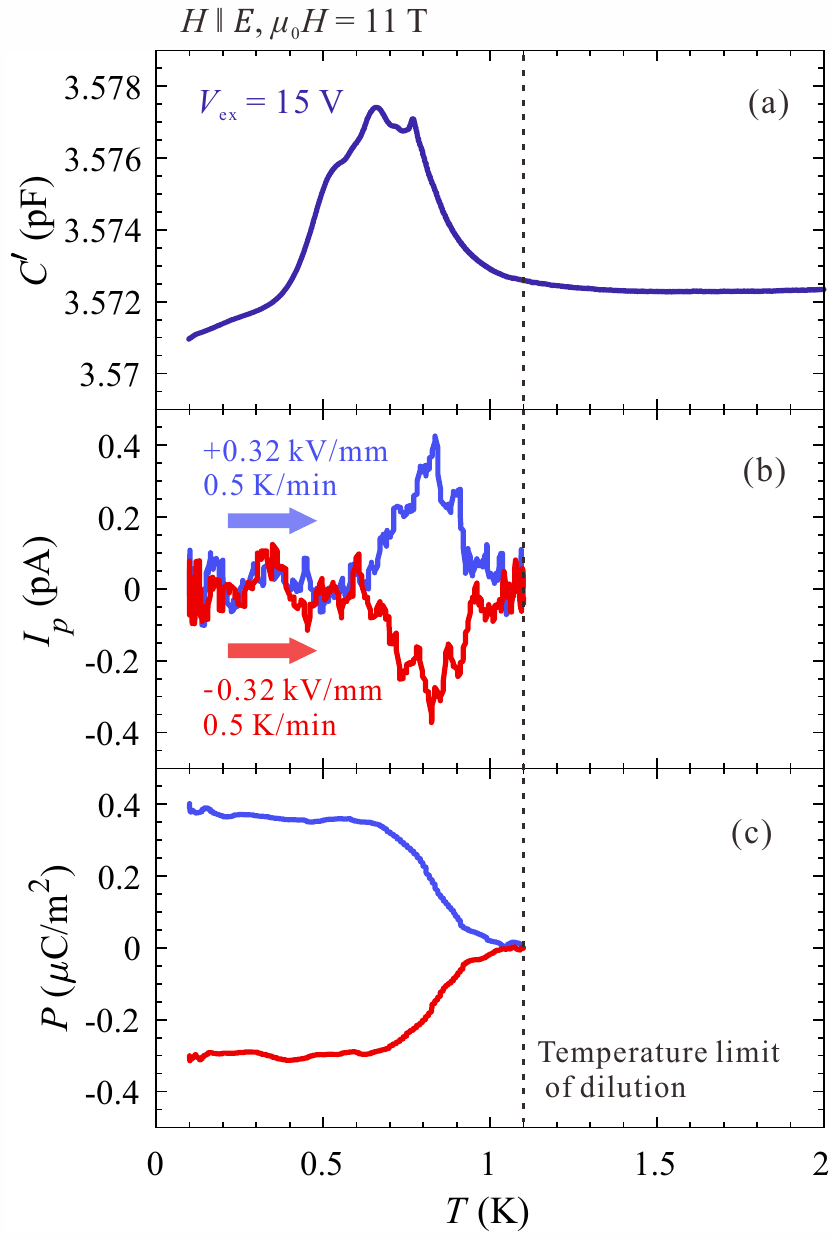}
\caption{Temperature scans of (a) capacitance $C'$ and (b) pyroelectric current $I_{p}$ measured in pelletized \RbCu sintered-powder samples at 11~T. The magnetic field is applied parallel to the electric field. (c) Electric polarization $P$ derived from the time integration of the pyroelectric current. The dashed line is the upper temperature limit of the $^3$He-$^4$He dilution refrigerator.}
\label{fig:polarization}
\end{figure}

Our first goal is to clarify the nature of the field-induced $T'$ dielectric anomaly.
In short, we find that it develops {\em only} in sintered powder but is absent in all samples derived from single crystals.
This is borne out in Fig.~\ref{fig:highT}, which shows the temperature dependence of sample-cell capacitance for samples of types (1)--(5).
The monotonic and featureless capacitance of the empty cell was measured separately and subtracted.
In an applied magnetic field, the capacitance of the {\em sintered-powder} sample develops a peak about 1.5~fF in the real part, as well as a distinct feature in the imaginary part at around $T'=8$~K [Fig.~\ref{fig:highT}(a)].
This behavior is very similar to the dielectric anomaly reported in Refs.~\cite{Yasui2013_2,Ueda2020} and clearly corresponds to the same ferroelectric phase transition.
However, the capacitance peak is completely absent in whole and powdered single crystals [see Figs.~\ref{fig:highT}(b)--\ref{fig:highT}(e)].
For the coaligned-crystals sample, the data in Fig.~\ref{fig:highT}(c) correspond to a configuration with $\mathbf{H}\bot\mathbf{b}$ and $\mathbf{H}\bot\mathbf{E}$.
Qualitatively similar data were obtained with $\mathbf{H}\bot\mathbf{b},\mathbf{H}\|\mathbf{E}$ and $\mathbf{H}\|\mathbf{b},\mathbf{H}\bot\mathbf{E}$.
The data for the single crystal in Fig.~\ref{fig:highT}(d) correspond to a magnetic field applied perpendicular to the $b$ axis.
The data taken for magnetic fields along the $b$ direction are qualitatively very similar.
We conclude that the $T'$ anomaly reported in Refs.~\cite{Yasui2013_1,Yasui2013_2,Reynolds2019,Ueda2020} {\em  is endemic to sintered powders and must be due to an impurity phase}.
Note that the data in Fig.~2(b) (random crystals) show a vague field-dependent feature of about 0.15~fF in the real capacitance, suggesting that there may be some residual impurity contamination of the large pile of tiny crystals that compose this sample.

The most likely culprit is a related copper molybdate, namely, Cu$_{3}$Mo$_{2}$O$_{9}$.
This material is known to have a huge ferroelectric anomaly at its antiferromagnetic transition, precisely at 8~K~\cite{Kuroe2011}.
The synthesis process for this compound~\cite{Hamasaki2008} is very similar to that in \RbCu.
In fact, we have identified isolated grains of this material in some of our crystal growth batches. To see how possible Cu$_{3}$Mo$_{2}$O$_{9}$ impurities might affect experiments on \RbCu, we measured the capacitance of a Cu$_{3}$Mo$_{2}$O$_{9}$ powder sample  using the same capacitor. The results are shown in Fig.~\ref{fig:highT}(f).
The anomaly at 8~K coincides with that seen in the sintered-powder sample of \RbCu.
Moreover, the magnitude of the $\Delta C'$ anomaly in Cu$_{3}$Mo$_{2}$O$_{9}$ is about 50 times larger.
As little as $\sim$2{\%} Cu$_{3}$Mo$_{2}$O$_{9}$ impurities would be enough to mimic the $T'$ anomaly in \RbCu sintered powders. Such a small amount would remain undetected by powder x-ray diffraction.
These results lead us to a clear conclusion: The $T'$ anomaly is not endemic to \RbCu.

Though not directly relevant to our main discussion, we make the following side note: In bulk single-crystal \RbCu we observe a different dielectric anomaly at about 19~K, as shown in Figs.~\ref{fig:highT}(b)--\ref{fig:highT}(d).
It is absent in the bulk material that is ground to a fine powder as shown in Fig.~\ref{fig:highT}(e).
In any case, this feature appears to be entirely field independent and is therefore of nonmagnetic origin.

\begin{figure*}[tbp]
\includegraphics[scale=1]{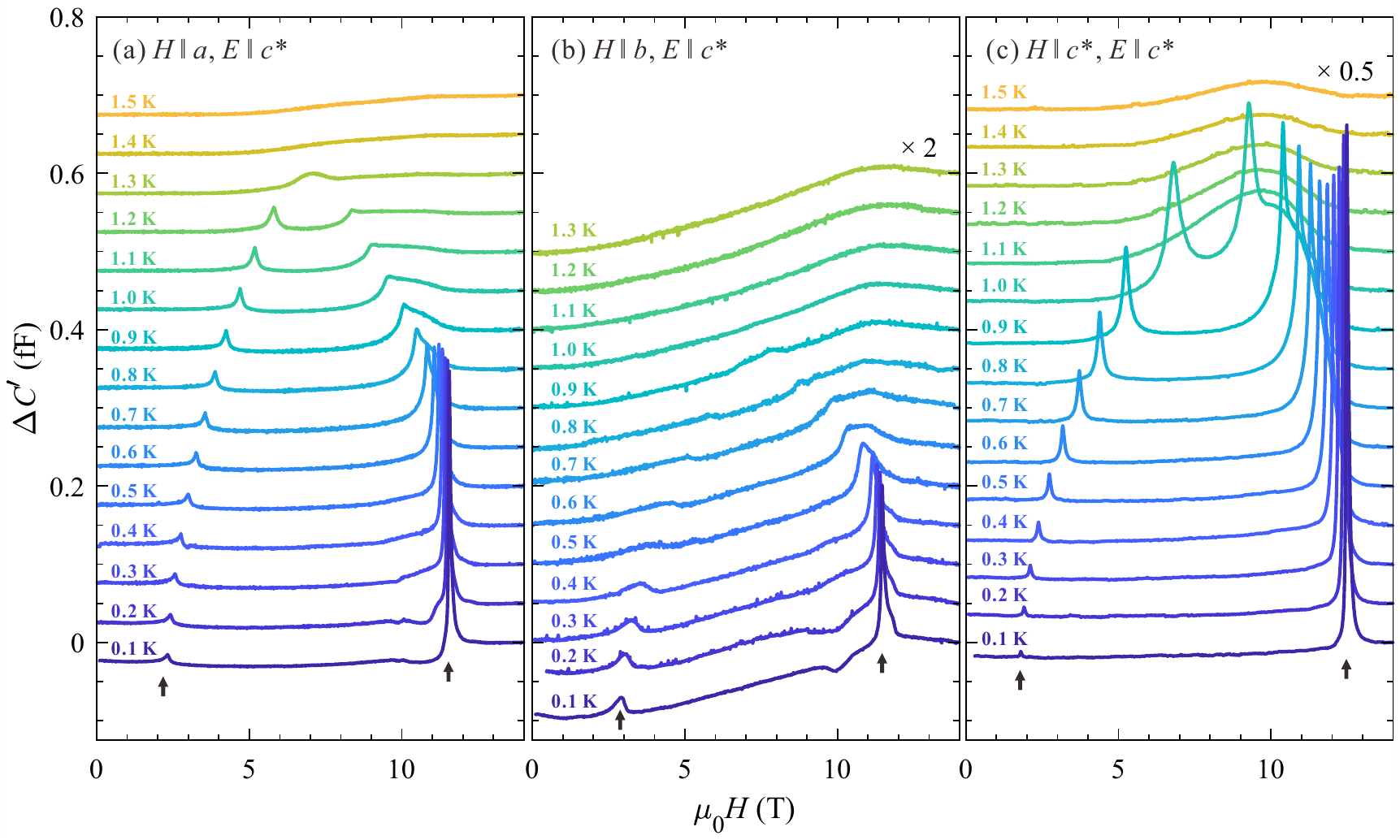}
\caption{Typical capacitance $\Delta C^{\prime}$ data in applied magnetic fields along the (a) $a$, (b) $b$, and (c) $c^{*}$ directions.
The label of the vertical axis is the change in the capacitance from 14-T $\Delta C'$.
For visibility, the scans are offset by 0.1~fF relative to one another.
Note that the data in (b) and (c) were previously rescaled by 2 and 0.5, respectively, in order to unify the scale among all plots.
Arrows indicate phase transitions as discussed in the text.}
\label{fig:rawCap}
\end{figure*}

\subsection{Low-temperature dielectric anomaly}
With the purported high-temperature chiral spin liquid phase and sample-related issues out of the way, we focus on the low-temperature behavior across the field-induced magnon BEC transition using the pelletized sintered-powder and single-crystal samples.

\subsubsection{Ferroelectric polarization}

First, we confirm that at precisely the field-induced  magnon BEC transition the sample indeed becomes ferroelectric.
Figure~\ref{fig:polarization} shows the temperature dependence of the capacitance, the pyroelectric current, and the corresponding electric polarization for the pelletized sintered-powder sample measured in a magnetic field $\mu_0H=11$~T.
In the capacitance, we clearly observe a broad peak at 0.7~K. It exactly corresponds to the phase boundary of the magnon BEC dome~\cite{Hayashida2019}, broadened due to random orientations of the powder grains relative to the applied-magnetic-field direction. This feature is consistent with the previous dielectric study below 6~T~\cite{Reynolds2019}.

Following the protocol described above, at $\mu_0H=11$~T we also clearly observe a pyroelectric current $I_{p}$ that peaks at the magnetic transition \footnote{In this measurement the actual sample temperature is estimated to lag about 0.1~K behind the read-out temperature due to the high heating rate}.
This current and the corresponding electrical polarization are completely switchable by the poling electric-field reversal [Figs.~\ref{fig:polarization}(b) and \ref{fig:polarization}(c)].
This tells us that spontaneous ferroelectric polarization is an inherent property of the magnon BEC phase.

As discussed in detail below, much higher quality and orientation-resolved data on the dielectric constant can be obtained in single crystals. Unfortunately, we were unable to also measure polarization in these samples. This is undoubtedly due to their tiny size. According to geometry considerations alone, the pyroelectric current expected in our single crystals is about 50 times smaller than that seen in the large pelletized powder sample, the latter already being at the limit of detectability using our setup.

\begin{figure}[tbp]
\includegraphics[scale=1]{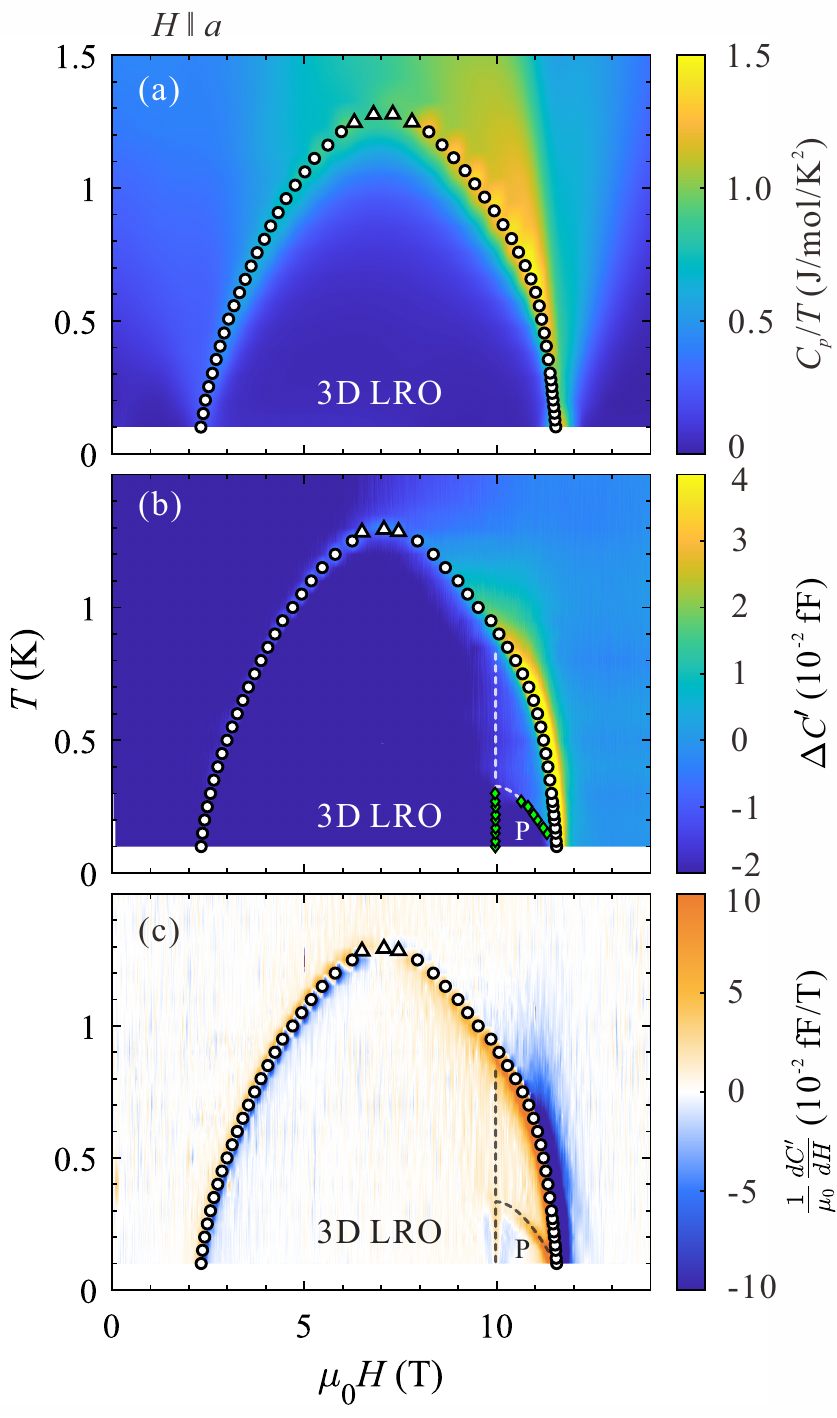}
\caption{False-color plots of (a) specific heat $C_{p}/T$, (b) capacitance $\Delta C'$, and (c) derivative of capacitance $\frac{1}{\mu_{0}}\frac{dC'}{dH}$ for $\mathbf{H}\| \mathbf{a}$.
Circles and triangles indicate the transitions identified in the field and temperature scans, respectively, for both specific heat and capacitance.
Green diamonds indicate extra features observed in the derivative of capacitance.
The phase regions are labeled as the three-dimensional long-range order (3D LRO) and presaturation (P).
Dashed lines are guides for the eye.}
\label{fig:PD_a}
\end{figure}

\begin{figure}[tbp]
\includegraphics[scale=1]{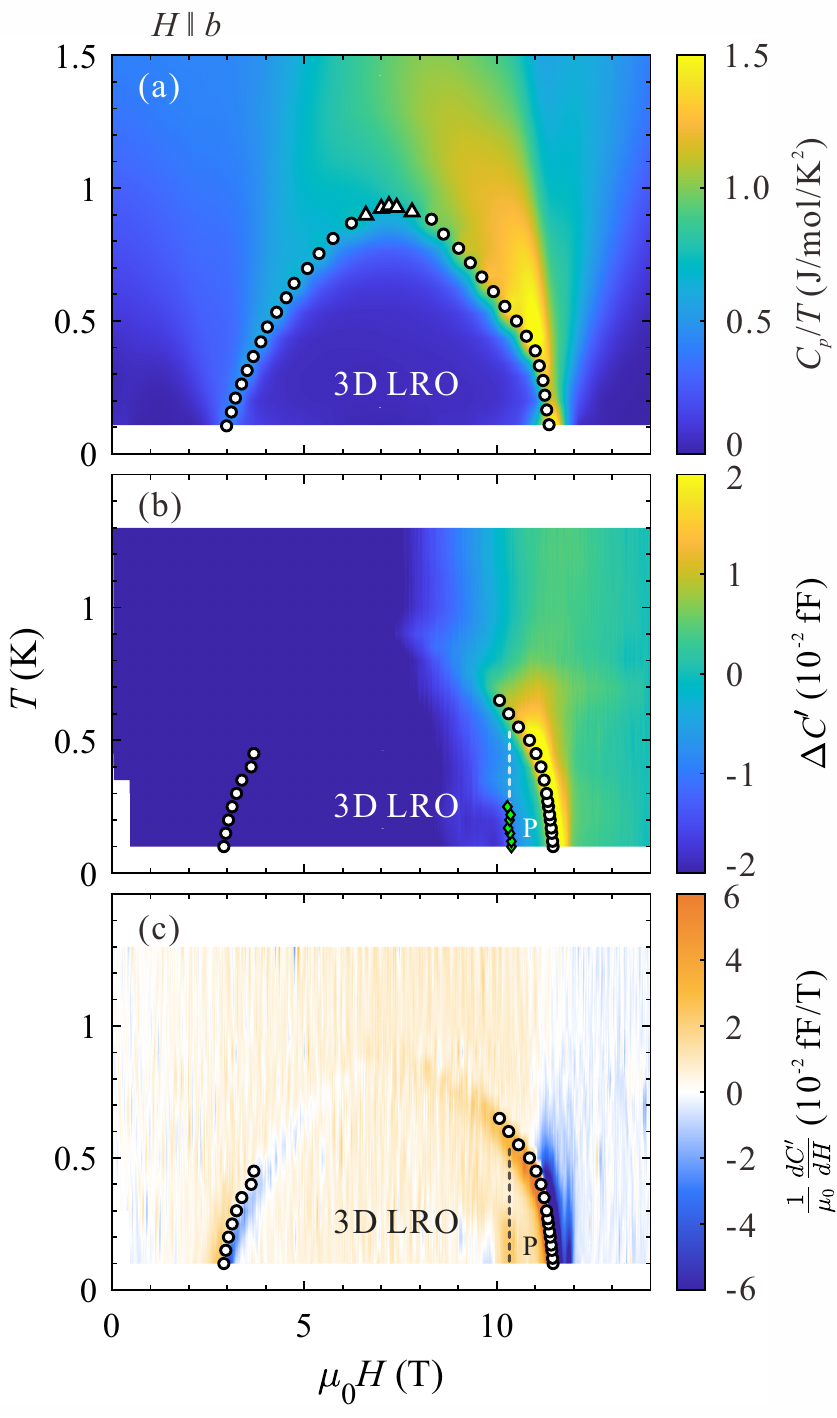}
\caption{False-color plots of (a) specific heat $C_{p}/T$, (b) capacitance $\Delta C'$, and (c) derivative of capacitance $\frac{1}{\mu_{0}}\frac{dC'}{dH}$ for $\mathbf{H}\| \mathbf{b}$.
Circles and triangles indicate the transitions identified in the field and temperature scans, respectively, for both specific heat and capacitance.
Green diamonds indicate extra features observed in the derivative of capacitance.
The phase regions are labeled as the three-dimensional long-range order (3D LRO) and presaturation (P).
Dashed lines are guides for the eye.}
\label{fig:PD_b}
\end{figure}

\begin{figure}[tbp]
\includegraphics[scale=1]{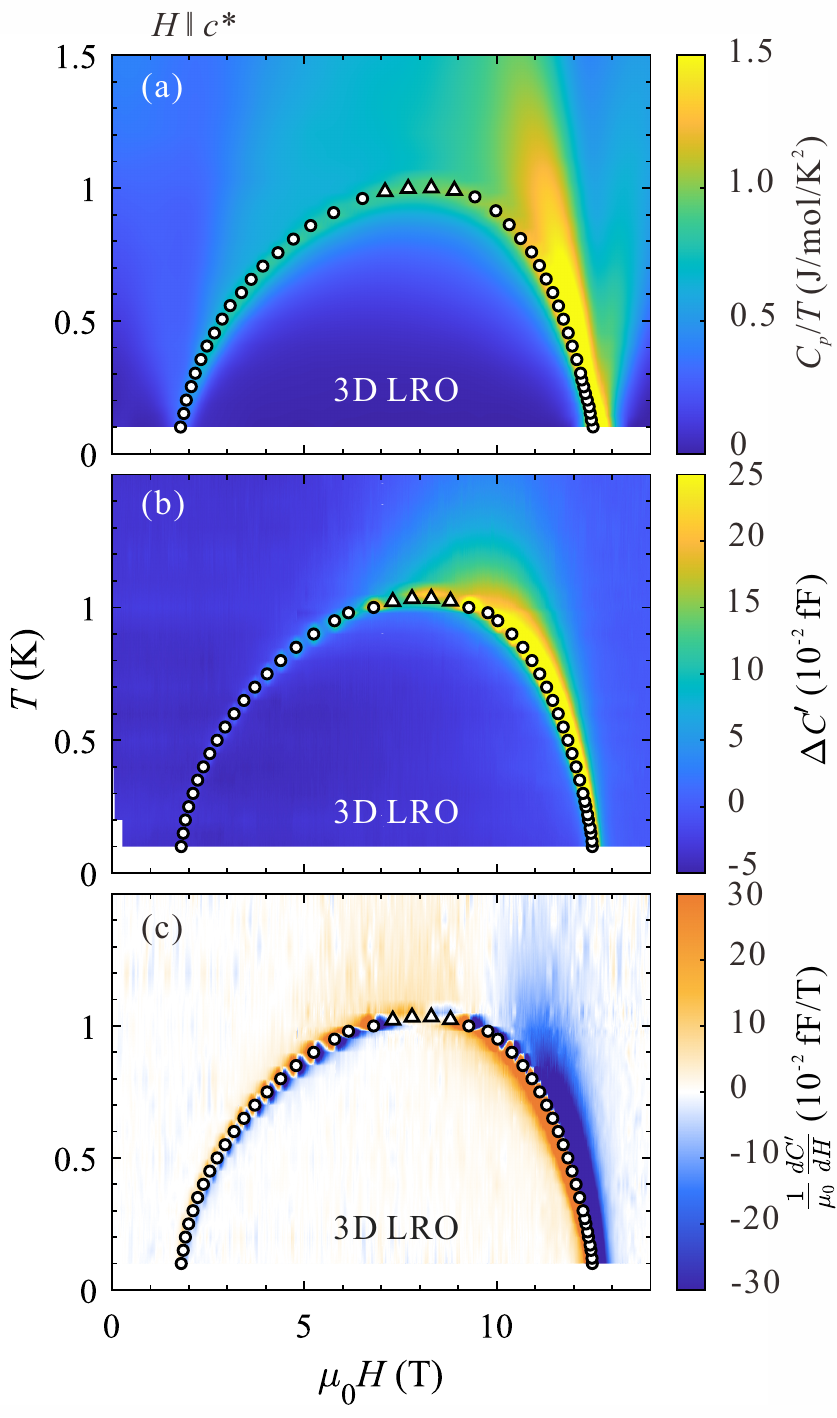}
\caption{False-color plots of (a) specific heat $C_{p}/T$, (b) capacitance $\Delta C'$, and (c) derivative of capacitance $\frac{1}{\mu_{0}}\frac{dC'}{dH}$ for $\mathbf{H}\| \mathbf{c}^{*}$.
Circles and triangles indicate the transitions identified in the field and temperature scans, respectively, for both specific heat and capacitance.
The three-dimensional long-range order phase (3D LRO) is labeled.
}
\label{fig:PD_c}
\end{figure}

\begin{figure}[tbp]
\includegraphics[scale=1]{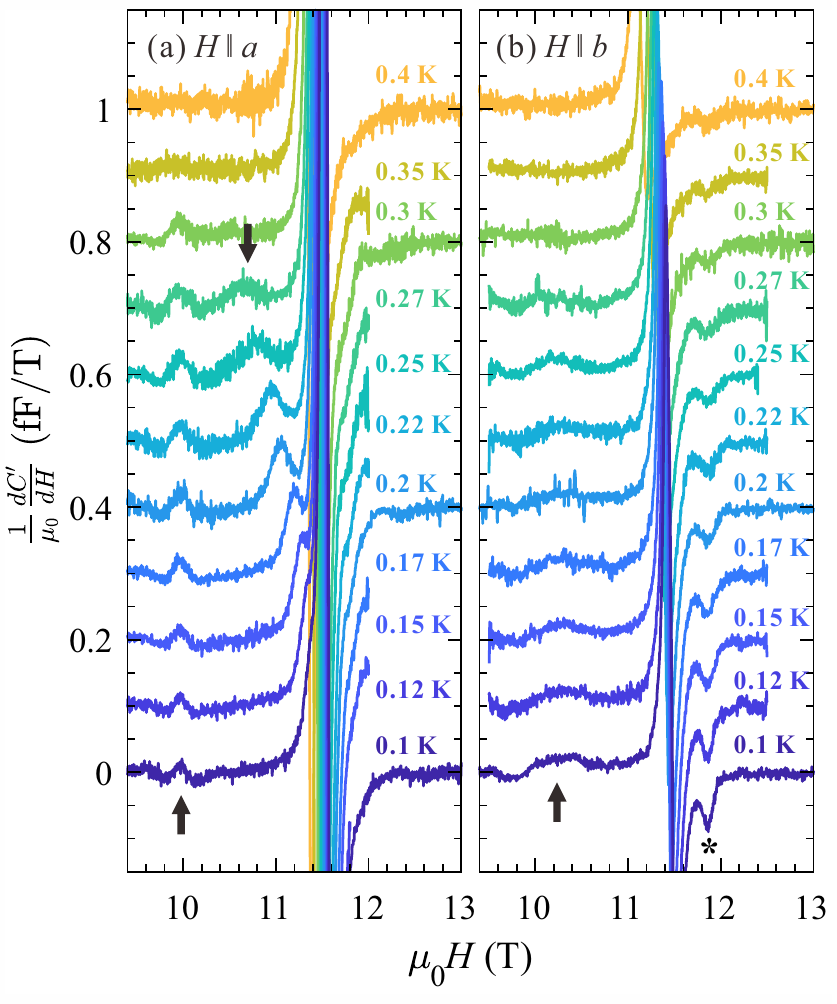}
\caption{Derivative of the capacitance for (a) $\mathbf{H}\|\mathbf{a}$ and (b) $\mathbf{H}\|\mathbf{b}$.
Arrows and an asterisk indicate additional phase transitions as discussed in the text.
}
\label{fig:dCdH}
\end{figure}

\subsubsection{Dielectric susceptibility}

In the spirit of scaling theory of phase transitions in which the free energy of a system is written as a sum of analytical (slowly varying) and singular (rapidly varying at the phase transition) parts,
we focus on the {\em critical} (diverging) contribution to dielectric susceptibility, which we henceforth denote as $\chi$. To separate this contribution out, we follow the approach of Ref.~\cite{Kim2014}. We assume that below $T=1$~K only this critical contribution shows any significant field dependence. We measure the capacitance $C$ of the sample ($C\approx 0.24$~pF corresponding to $\epsilon \sim 10\epsilon_{0}$) and subtract a baseline value measured at 14~T (well outside the critical regime) at the same temperature: $\Delta C = C(H)-C(14~{\rm T})$. 
In our experiments, $\Delta C/C\lesssim 1\%$. We further assume that the critical susceptibility $\chi$ is proportional to $\Delta C$. 
This approach is, admittedly, not without limitation, as a separation between analytical and singular free energy contributions is not unambiguous to begin with. Indeed, the measured $\Delta C$ shows some small variation between the gapped, ordered, and fully polarized phases even away from the critical regions of the field-induced phase transitions. 
As will be discussed below, at least for the $\mathbf{H}\|\mathbf{c}^{*}$ geometry central for this study, this effect is considerably smaller than the divergence in the immediate vicinity of the actual critical point.

Typical raw data for the three field configurations are shown in Fig.~\ref{fig:rawCap}.
Pairs of sharp peaks are clearly observed below 1.4~K for $\mathbf{H}\|\mathbf{a}$, 0.9~K for $\mathbf{H}\|\mathbf{b}$, and 1~K for $\mathbf{H}\|\mathbf{c}^{*}$, correspondingly.
Only the real part $\Delta C'$ of the measured capacitance is plotted. 
 An imaginary contribution is also present and will be noted in Sec.~\ref{scaling}.
These anomalies are very similar to those found at ferroelectric transitions in other magnon BEC compounds such as \TCL~\cite{Kimura2016,Kimura2017}. The only difference is that our measurements on \RbCu extend to higher fields closer to saturation.

Since in our experiments the capacitor plates are directly deposited on the crystal surface, associating $\chi$ with $\Delta C$ relies on the assumption that striction effects are negligible. 
To check that, we also measured the capacitance of the above-mentioned coaligned single crystals loaded in a capacitance cell [Fig.~\ref{fig:samples}(c)]. 
In this setup the crystals are in contact with only the lower plate, while the upper plate remains free. The thus measured capacitance is entirely insensitive to striction and is only affected by changes in the sample's dielectric constant.
The susceptibility divergence using that setup is practically identical to that measured with electrodes mounted directly on the sample surface [Fig.~\ref{fig:samples}(d)], validating our approach.

The bulk of the collected data are visualized in false-color plots in Figs.~\ref{fig:PD_a}(b), \ref{fig:PD_b}(b), and \ref{fig:PD_c}(b).
The peak positions are best determined in analyzing the $dC'/dH$ curves measured at a constant temperature.
That quantity is visualized in Figs.~\ref{fig:PD_a}(c), \ref{fig:PD_b}(c), and \ref{fig:PD_c}(c).
Open circles in the (b) and (c) panels correspond to points where the derivative changes sign.

For a direct comparison of the dielectric and calorimetric responses we combine  the $\mathbf{H}\|\mathbf{b}$ specific heat data of  Ref.~\cite{Hayashida2019} with additional measurements for $\mathbf{H}\|\mathbf{a}$ and for $\mathbf{H}\|\mathbf{c}^{*}$ under identical conditions, as shown in false-color plots in Figs.~\ref{fig:PD_a}(a), \ref{fig:PD_b}(a), and \ref{fig:PD_c}(a).
Open circles mark the observed lambda anomalies.
The latter coincide with the observed dielectric divergences and follow the same anisotropy pattern with regard to the magnetic-field direction.
We conclude that the divergent dielectric constant is associated with the thermodynamic field-induced phase transition and the onset of magnetic order.

Notably, the dielectric signal is most pronounced for the $\mathbf{H},\mathbf{E}\|\mathbf{c}^{*}$ configuration. This observation allows us to exclude one possible microscopic origin of ferroelectricity in \RbCu, namely, the reverse Dzyaloshinskii-Moriya mechanism for spiral magnetic order~\cite{Katsura2005,Mostovoy2006}. In that model the key ingredient is a planar helimagnetic structure with electric polarization appearing in the spin-rotation plane. A planar helix orients itself to be perpendicular to an applied magnetic field, excluding ferroelectricity in the $\mathbf{H}\|\mathbf{E}$ setting.

%%Presaturation phase
Weak features in the dielectric response [Figs.~\ref{fig:PD_a}(b), \ref{fig:PD_a}(c), \ref{fig:PD_b}(b), and \ref{fig:PD_b}(c)] reveal additional phase transitions near the upper critical fields for $\mathbf{H}\|\mathbf{a}$ and $\mathbf{H}\|\mathbf{b}$ which are overlooked by specific heat experiments.
The tentative boundaries of the phases are sketched with dashed lines and small green diamonds.  
The corresponding features are best seen in the $dC'/dH$ curves in Fig.~\ref{fig:dCdH} as highlighted by the arrows.
For $\mathbf{H}\|\mathbf{b}$ there seems to be an additional feature at 11.9~T, above the previously established saturation field~\cite{Hayashida2019}, as indicated by an asterisk in Fig.~\ref{fig:dCdH}(b). Although we cannot exclude it being a new phase transition, this is unlikely.
The most feasible interpretation is a small misaligned crystalline grain in the sample, for example, in the $\mathbf{H}\|\mathbf{c}^{*}$ orientation. While the peak for $\mathbf{H}\|\mathbf{b}$ is the weakest of all the orientations [Fig.~\ref{fig:rawCap}(b)], the peak for $\mathbf{H}\|\mathbf{c}^{*}$ is very strong at the saturation field, which is higher than 11.4~T [Fig.~\ref{fig:rawCap}(c)]. Consequently, even a tiny grain could be responsible for the small dip at 11.9~T for $\mathbf{H}\|\mathbf{b}$.
On the other hand, any features below the saturation field cannot be due to a similar contamination, since other crystal orientations show no particularly strong features in that field range.

The microscopic nature of the presaturation phases cannot be further clarified at this point.
That said, spin-nematic~\cite{Hikihara2008,Sudan2009} or spin density wave~\cite{Sato2013} presaturation phases are only to be expected in a frustrated ferro-antiferromagnet such as \RbCu, particularly due to its proximity~\cite{Ueda2020} to the classical ferromagnet-helimagnet transition~\cite{Ueda2013,Starykh2014}.

\subsubsection{Choice of excitation voltage and frequency}
\begin{figure}[tbp]
\includegraphics[scale=1]{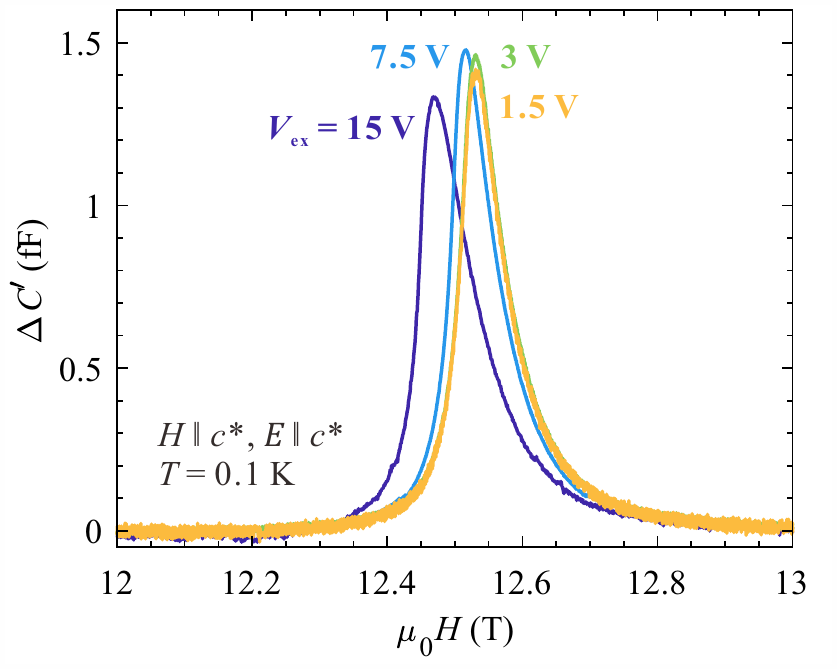}
\caption{Field scans of $\Delta C'$ at 0.1~K with a series of excitation voltages for $\mathbf{H}\|\mathbf{c}^{*}$.
}
\label{fig:Vex}
\end{figure}

\begin{figure}[tbp]
\includegraphics[scale=1]{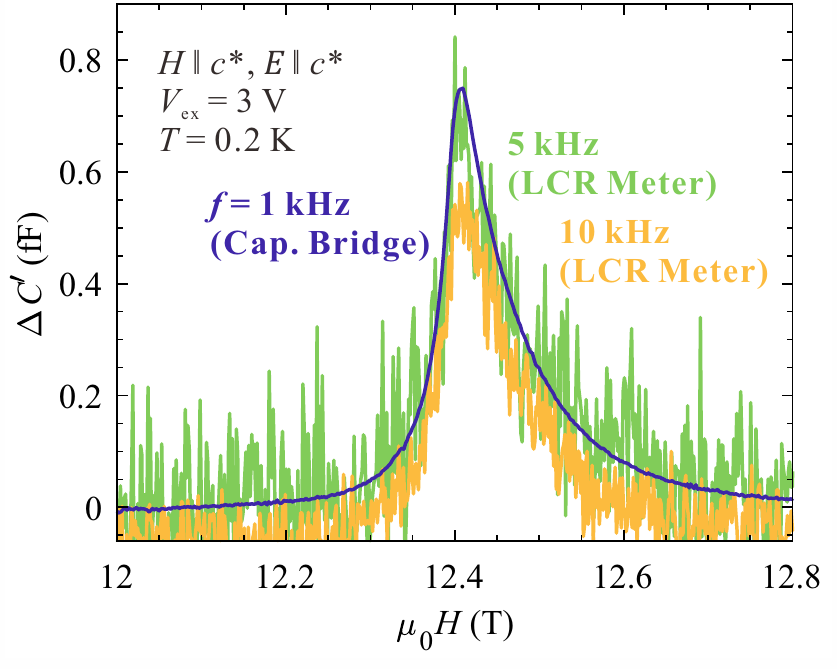}
\caption{Frequency variation of field scans of $\Delta C'$ at 0.2~K measured with an excitation voltage of $V_{\rm ex}=3$~V for $\mathbf{H}\|\mathbf{c}^{*}$. Cpa., capacitance.
}
\label{fig:frequency}
\end{figure}

\subsection{BEC quantum criticality}

%Bias and amplitude tests
Our ultimate goal is to capture the critical susceptibility in the field-induced magnon BEC QCP via the dielectric channel.
Measuring critical behavior is always a delicate matter.
To avoid common mistakes we started with carrying out a few tests and establishing measurement and data treatment procedures.

A potential pitfall to avoid is selecting an inappropriately large excitation voltage $V_{\rm ex}$ for alternating current (ac) permittivity measurements.
Figure~\ref{fig:Vex} shows data collected at the upper critical field  in $\mathbf{H}\|\mathbf{c}^{*}$ at the lowest attainable temperature $T=0.1$~K.
We see that the permittivity peak shifts to slightly higher fields upon reducing the excitation voltage from 15 to 7.5~V.
This is likely due to a slight heating of the sample by the probing ac field, which in turn originates from a small dissipative component to capacitance, as noted below in more detail.
The unwanted heating is completely gone for $V_{\rm ex}\lesssim3$~V.
Indeed, an excitation voltage of 1.5~V produces practically identical data to those of 3~V.

Another concern is whether the selected frequency is low enough to probe the static limit of susceptibility. To check this, we
studied the frequency dependence of the capacitance peak at the upper critical field at 0.2~K for $\mathbf{H}\|\mathbf{c}^{*}$ (Fig.~\ref{fig:frequency}).
Since the capacitance bridge operates only at 1~kHz, other frequencies were probed with a less sensitive {\it LCR} meter.
Noise in the latter setup not withstanding, there is virtually no change in either peak shape or amplitude  between 1 and 5~kHz. 
There are indeed some small yet discernible deviations at 10~kHz, possibly due to sample heating, but a frequency of 1~kHz can clearly be viewed as ``safe.''

All data discussed below were collected with $V_{\rm ex}=3$~V and $f=1$~kHz.

\subsubsection{Is polarization the primary order parameter?}

Next we question  whether electric polarization $\mathbf{P}$ is indeed a {\em primary} order parameter in the {\em quantum} phase transition at $T\rightarrow 0$. This is not a trivial issue.
Similar to our case, early studies of the field-induced magnon BEC transition in \SUL detected a strong sharp peak in the dielectric constant~\cite{Schrettle2013}.
It was later shown that at low temperatures the divergence decreases drastically.
Eventually at the QCP (in the $T\rightarrow 0$ limit) the peak is reduced to a finite small jump at the transition point~\cite{Povarov2015}.
At the QCP, electric polarization is merely a {\em secondary} order parameter.
This scenario repeats in \TCL~\cite{Kimura2016,Kimura2017}: A prominent dielectric divergence at finite temperature gives way to a vanishingly small anomaly at the QCP.
Such behavior is easily explained. 
In both systems, ferroelectricity is due to the reverse Dzyaloshinskii-Moriya mechanism~\cite{Katsura2005,Mostovoy2006}.
In the free energy it linearly couples to the vector chirality $\langle \mathbf{S}_1\times \mathbf{S}_2\rangle$ of the magnetic subsystem.
The latter is proportional to the product of the field-induced uniform longitudinal magnetization $\langle S_z\rangle$  and the spontaneous transverse magnetization (BEC order parameter) $\langle S_\bot\rangle$.
At $T=0$, due to the spin gap, $\langle S_z\rangle$ is strictly zero all the way up to $H_{c1}$ and grows linearly above it~\cite{Matsumoto2002}.
As a result, precisely at the QCP the bilinear coupling of polarization to the magnetic order parameter vanishes.
Only higher-order (quadratic) coupling remains relevant, making $\mathbf{P}$  a secondary order parameter of the phase transition. 
Dielectric susceptibility only shows a finite jump at $H_{c1}$ with no divergence.
The situation changes at a finite temperature.
Here, the thermalized quantum paramagnet has a nonzero magnetization $\langle S_z\rangle$ induced by the external field.
Vector chirality becomes proportional to  $\langle S_\bot\rangle$.
Electric polarization becomes linearly coupled to the primary order parameter and turns itself into one.

As is evident from Fig.~\ref{fig:rawCap}, the same scenario plays out in \RbCu at the {\em lower} critical field.
A sharp dielectric divergence at finite temperature weakens and almost vanishes at low temperatures.
At the {\em quantum} phase transition, the dielectric constant is not critical.
In contrast, at the {\em upper} critical field the anomaly only becomes {\em stronger} upon cooling and persists at the QCP. This behavior is just as easily explained.
Near $H_{c2}$ the system has a large uniform magnetization even at $T=0$.
Vector chirality is proportional to $\langle S_\bot\rangle$ with a large prefactor $\langle S_z\rangle\sim 1/2$.
Polarization remains the primary order parameter {\em even at the QCP}.
A similar argumentation holds for the spin-dependent $p$-$d$ hybridization mechanism~\cite{Kim2014}.
In the discussion, we shall focus our attention on the upper critical field in \RbCu.

\begin{figure}[tbp]
\includegraphics[scale=1]{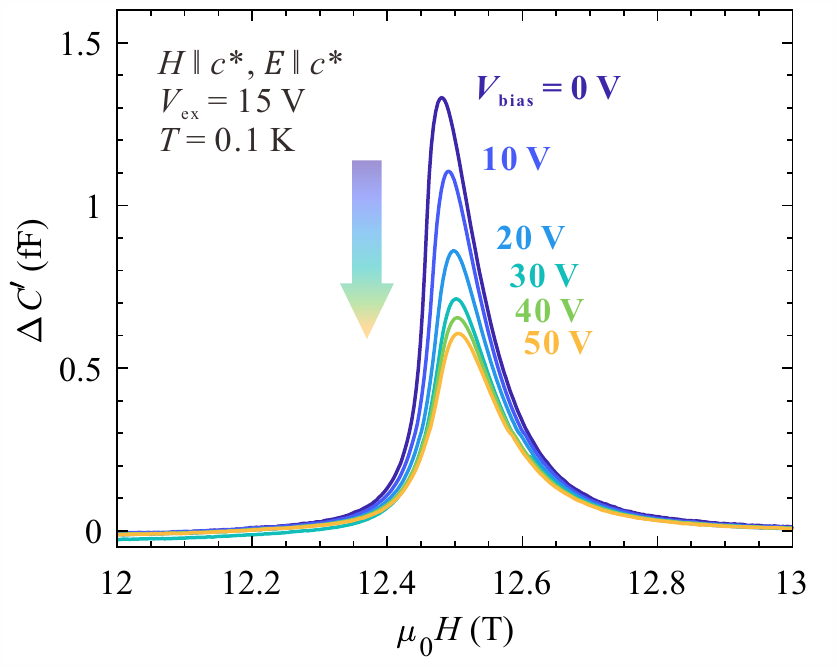}
\caption{Field scans of $\Delta C'$ at 0.1~K with a series of bias dc voltages for $\mathbf{H}\|\mathbf{c}^{*}$ measured with an excitation voltage of $V_{\rm ex}=15$~V.
}
\label{fig:Vbias}
\end{figure}

Whether or not polarization is a primary order parameter can be directly tested by studying the effect of an external electric field.
Applying a field conjugate to a {\em secondary} order parameter  may at most shift the transition point, but otherwise will leave the transition and the susceptibility divergence intact.
A field conjugate to a {\em primary} order parameter breaks the transition symmetry externally. It thereby destroys the transition, suppressing the critical divergence of susceptibility.
Figure~\ref{fig:Vbias} shows the bias dc voltage dependence of the permittivity peak measured in \RbCu at $H_{c2}$ in $\mathbf{H}\|\mathbf{c}^{*}$ at $T=0.1$~K.
The peak is clearly suppressed by the applied electric field, confirming electric polarization as a primary order parameter.

\subsubsection{BEC universality class?}
\begin{figure}[tbp]
\includegraphics[scale=1]{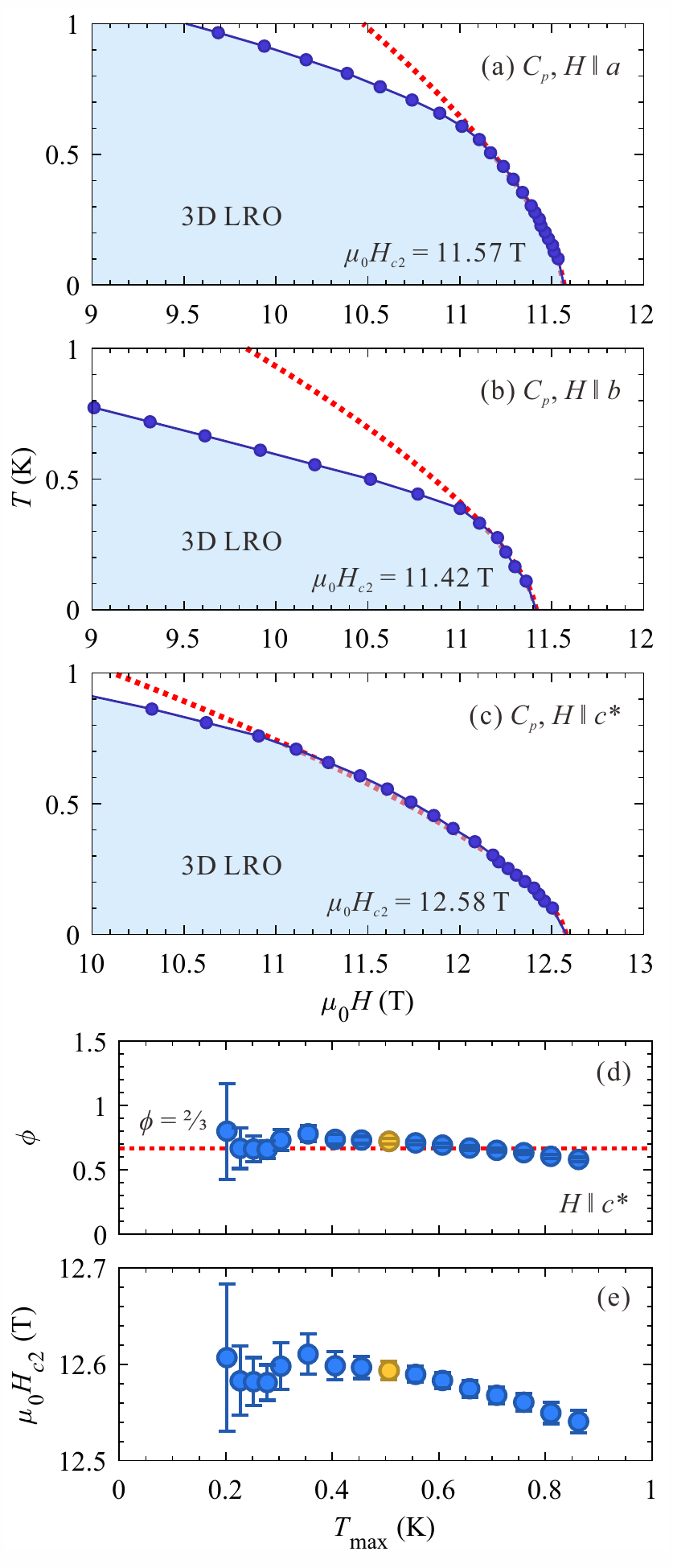}
\caption{Phase diagram boundaries measured by specific heat in applied magnetic fields along the (a) $a$, (b) $b$, and (c) $c^{*}$ directions.
The dashed curves are power-law fits in the range of $T\leq 0.35$~K, where the exponent is fixed to $2/3$.
(d) and (e) Shrinking-fit-window analysis of the measured phase boundaries for $\mathbf{H}\|\mathbf{c}^{*}$.
The plots show the least-squares fitted values of (d) the  phase boundary exponent $\phi$ and (e) the upper critical field $H_{c2}$ vs the temperature range used for the fit. The dashed line is at $\phi=2/3$.
}
\label{fig:HC_boundary}
\end{figure}

\begin{figure}[tbp]
\includegraphics[scale=1]{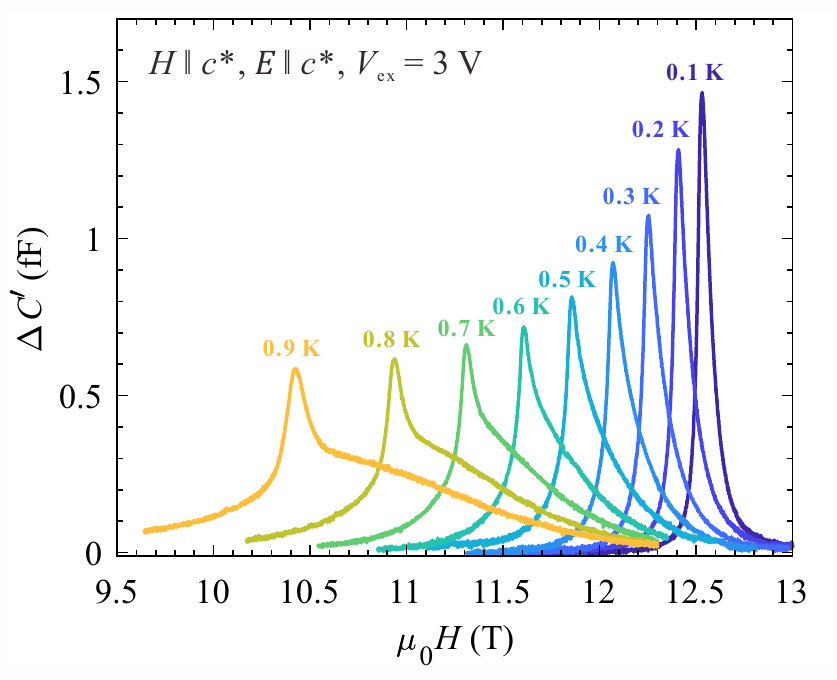}
\caption{Field scans of $\Delta C'$ measured with an excitation voltage of $V_{\rm ex}=3$~V for $\mathbf{H}\|\mathbf{c}^{*}$.
}
\label{fig:constT}
\end{figure} 
Another issue is whether the field-induced transitions in \RbCu can indeed be approximated as being in the BEC universality class.
The main concern is the magnetic anisotropy that is very apparent in the geometry-dependent shapes of the ``domes'' of magnetic long-range order in the measured phase diagrams.
To elucidate its effect on the nature of the quantum critical point at $H_{c2}$, in Figs.~\ref{fig:HC_boundary}(a)--\ref{fig:HC_boundary}(c), we plot the details of the magnetic phase boundary in that field region, as deduced from the lambda anomalies in the specific heat measurements.
These boundaries carry information on the so-called crossover exponent $\phi$, which is defined as
\begin{equation}
\left| H - H_{c}\right| \propto T^{1/\phi}
\end{equation}
in the limit  $T\rightarrow 0$.
For a 3D BEC QCP, we expect $\phi=2/3$~\cite{Nikuni2000,Giamarchi2008,Zapf2014}.
In the case of \RbCu, for $\mathbf{H}\|\mathbf{a}$ and $\mathbf{H}\|\mathbf{b}$ there is a clear change of slope in the curve indicating a crossover between scaling regimes.
This may or may not be related to the presence of presaturation phases in these field configurations.
Regardless, the  regime close to the QCP is too narrow for any quantitative scaling analysis.
These geometries are clearly not suitable for our study of BEC criticality.
No anomalous change in slope is seen for $\mathbf{H}\|\mathbf{c}^{*}$.
As shown in Figs.~\ref{fig:HC_boundary}(d) and \ref{fig:HC_boundary}(e), a windowing analysis~\cite{Huvonen2012} of the power-law fit to the phase boundary yields a crossover exponent $\phi=0.72(2)$.
In fact, the $\phi=2/3$ BEC values describe the data very well up to 0.7~K or in a 1.5-T range below the transition field.
For this particular geometry and those temperatures accessible in our experiments, the transition does indeed look very much like a BEC one.

\begin{figure}[tbp]
\includegraphics[scale=1]{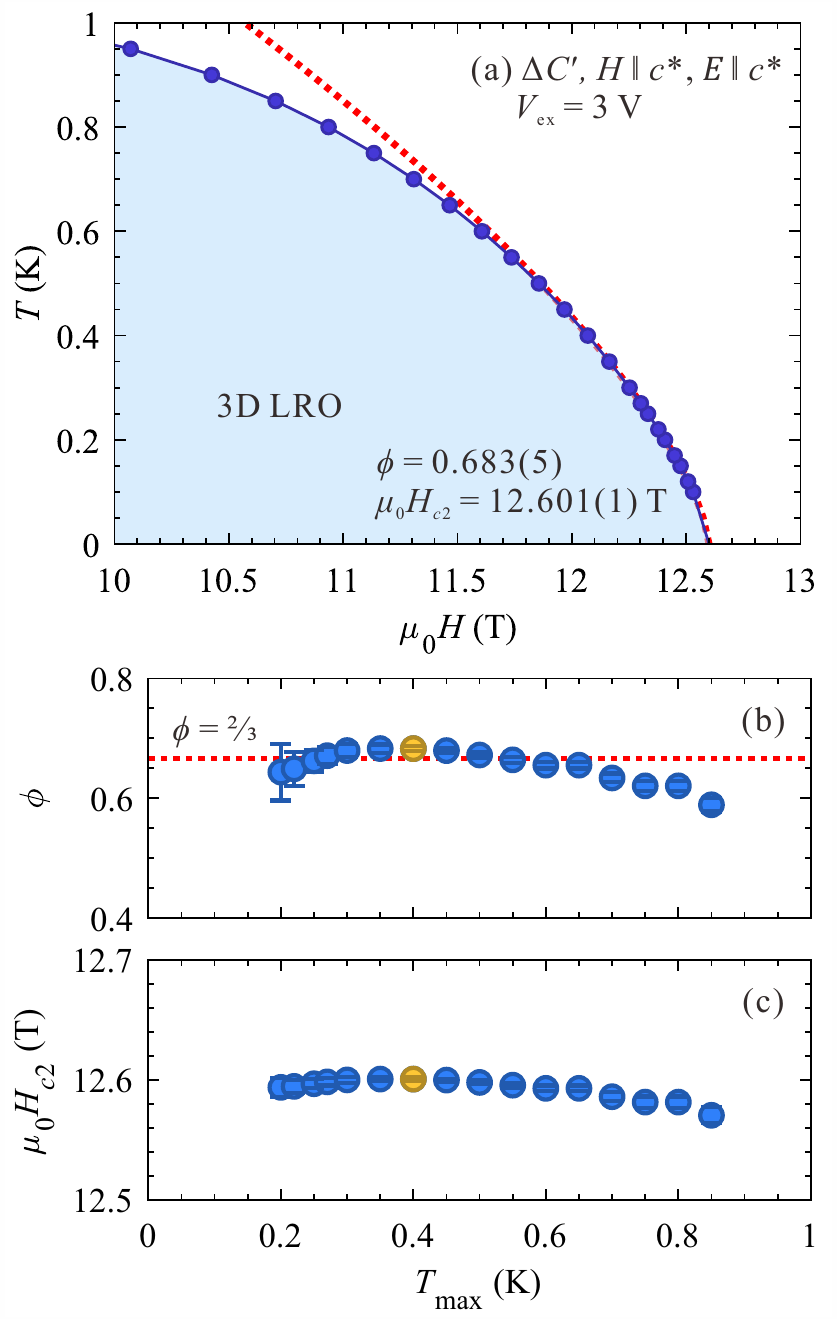}
\caption{(a) Phase diagram boundary for $\mathbf{H}\|\mathbf{c}^{*}$. Circles are the transition points determined from the capacitance data. The solid curve is a guide for the eye. The dashed curve is a power-law fit in the range of $T\leq 0.4$~K, where the best results were obtained.
(b) and (c) Windowing analysis of the measured phase boundaries. The plots show the least-squares fitted values of (b) the phase boundary exponent $\phi$ and (c) the upper critical field $H_{c2}$ vs the temperature range used for the fit. The dashed line is at $\phi=2/3$.
}
\label{fig:phase}
\end{figure}

The same conclusion is reached by analyzing the $\mathbf{H}\|\mathbf{c}^{*}$ phase boundary determined by the positions of the dielectric peaks.
Typical constant-$T$ capacitance scans measured with $V_{\mathrm{ex}}=3$~V are shown in Fig.~\ref{fig:constT}, and the resulting phase boundary is shown in Fig.~\ref{fig:phase}(a).
The windowing analysis shown in Figs.~\ref{fig:phase}(b) and \ref{fig:phase}(c) reveals an even more robust agreement with BEC behavior, yielding $\phi=0.683(5)$ and $\mu_{0}H_{c2}=12.601(1)$~T for the optimal temperature fitting range $T_{\mathrm{max}}=0.4$~K.
This is actually not surprising, since the dielectric anomaly is strongest in the all-important low-$T$ region, while the specific heat lambda anomaly becomes very small and hard to measure (Nernst's theorem).

\begin{figure}[tbp]
\includegraphics[scale=1]{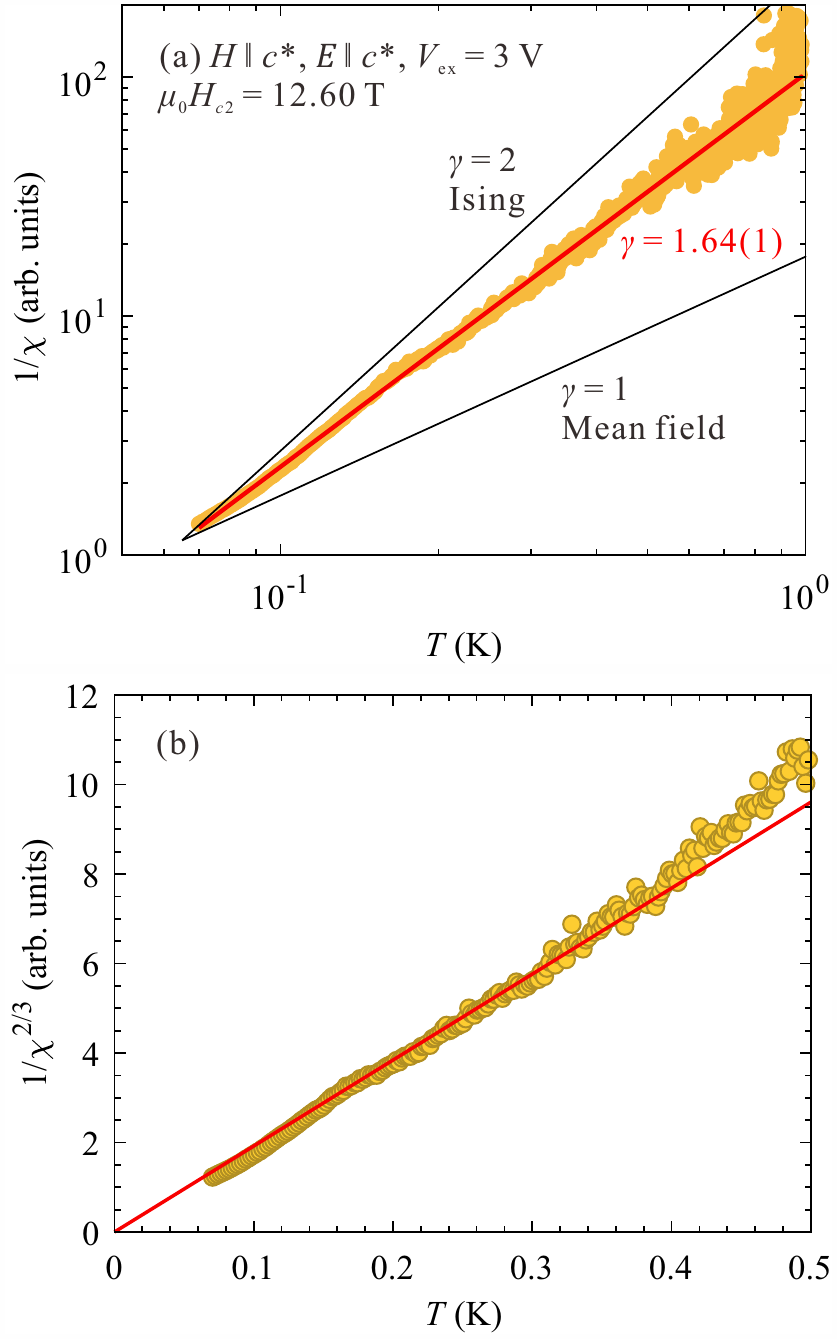}
\caption{(a) A log-log plot of the inverse critical dielectric susceptibility, $1/\chi$. The red line is a power-law fit to the data. Black lines are power laws with exponents of 1 and 2. (b) Inverse critical dielectric susceptibility $1/\chi^{2/3}$ vs temperature $T$ at $H=H_{c2}$. The data are binned by two successive points. The solid line is a guide for the eye.}
\label{fig:QCP}
\end{figure}

\subsubsection{Scaling of susceptibility at QCP}\label{scaling}
Having selected an appropriate geometry, we finally proceed to examine the dielectric susceptibility measured along the quantum critical trajectory $\mu_{0}H=\mu_{0}H_{c2}$ ($=12.60$~T) for $\mathbf{H}\|\mathbf{c}^{*}$.
In these measurements the imaginary part of the capacitance is below $0.1${\%} of the real one and never exceeds 11{\%} of $\Delta C'$. We therefore ignore the the imaginary part and analyze the scaling of the real part alone. 
Furthermore, we revisit the issue of the slight field dependence of the measured capacitance outside the critical regions. 
The ``background'' capacitance seems to be rather flat in the gapped phase, steadily increases in the ordered phase, and again becomes flat in the fully polarized state [Fig.~\ref{fig:rawCap}(c)], with small additional steps at each transition. 
Magnetostriction is one possible cause for this behavior. Regardless, for the selected $\mathbf{H}\|\mathbf{c}^{*}$ geometry, the entire effect is almost two orders of magnitude smaller than the actual divergence. 
It is minimized even further by the fact that the critical trajectory lies entirely in the paramagnetic state and $\Delta C$ is defined through a  subtraction of the baseline value measured in the same phase.

We then plot $1/\chi \propto1/\Delta C'$ vs temperature in a log-log plot as shown by the symbols in Fig.~\ref{fig:QCP}(a).
A power-law behavior is apparent, confirming that the critical contribution is dominant in the extracted $\chi$.
A least-squares fit by a power law yields an exponent $\gamma=1.64(1)$.
This is substantially distinct from the $\gamma=1$ divergence at the mean-field transition and the Ising exponent $\gamma=2$ observed in  Ba$_{2}$CoGe$_{2}$O$_{7}$~\cite{Kim2014}.
On the other hand, the observed exponent is in excellent agreement with $\gamma=3/2$ expected for a three-dimensional BEC quantum critical susceptibility~\cite{Kim2014,Continentino2011}.
That the dielectric susceptibility indeed diverges as $T^{3/2}$ along the critical trajectory as predicted for BEC is even more visible in Fig.~\ref{fig:QCP}(b). The solid line is not a fit, but simply a guide for the eye.

\section{Conclusion}
\RbCu is an amazing quantum multiferroic, but only single crystal experiments can provide a complete picture of its magnetoelectric properties.
We make the following conclusions: (i) The dielectric anomaly at 8~K reported in Refs.~\cite{Yasui2013_1,Yasui2013_2,Reynolds2019,Ueda2020}, ascribed to a chiral spin liquid phase~\cite{Furukawa2010}, is actually due to a tiny amount of a ferroelectric impurity.
 (ii) \RbCu is a rare species with experimentally accessible field-induced magnon BEC transitions at which the electric polarization is a primary order parameter at the quantum critical point.
In any gapped quantum paramagnet and regardless of the microscopic mechanism of magnetoelectric coupling, this situation cannot occur at the lower critical field.
(iii) The bilinear coupling of electric polarization and the magnetic BEC order parameter enables breakthrough direct measurements of the critical BEC susceptibility.
Its measured power-law divergence at the quantum critical point is in excellent agreement with long-standing theoretical predictions.
(iv) A presaturation phase is discovered and may represent an exotic order such as a spin-nematic state or spin density wave.
This illustrates that electrometry can be an extremely sensitive experimental tool for quantum magnetism.
 
\section*{Acknowledgements}
We thank Dr. D.~Blosser for assistance with the dielectric measurements.
This work was supported by the Swiss National Science Foundation, Division II.
 
%\bibliography{Rb2Cu2Mo3O12_Dielectric}

\begin{thebibliography}{52}%
\makeatletter
\providecommand \@ifxundefined [1]{%
 \@ifx{#1\undefined}
}%
\providecommand \@ifnum [1]{%
 \ifnum #1\expandafter \@firstoftwo
 \else \expandafter \@secondoftwo
 \fi
}%
\providecommand \@ifx [1]{%
 \ifx #1\expandafter \@firstoftwo
 \else \expandafter \@secondoftwo
 \fi
}%
\providecommand \natexlab [1]{#1}%
\providecommand \enquote  [1]{``#1''}%
\providecommand \bibnamefont  [1]{#1}%
\providecommand \bibfnamefont [1]{#1}%
\providecommand \citenamefont [1]{#1}%
\providecommand \href@noop [0]{\@secondoftwo}%
\providecommand \href [0]{\begingroup \@sanitize@url \@href}%
\providecommand \@href[1]{\@@startlink{#1}\@@href}%
\providecommand \@@href[1]{\endgroup#1\@@endlink}%
\providecommand \@sanitize@url [0]{\catcode `\\12\catcode `\$12\catcode
  `\&12\catcode `\#12\catcode `\^12\catcode `\_12\catcode `\%12\relax}%
\providecommand \@@startlink[1]{}%
\providecommand \@@endlink[0]{}%
\providecommand \url  [0]{\begingroup\@sanitize@url \@url }%
\providecommand \@url [1]{\endgroup\@href {#1}{\urlprefix }}%
\providecommand \urlprefix  [0]{URL }%
\providecommand \Eprint [0]{\href }%
\providecommand \doibase [0]{https://doi.org/}%
\providecommand \selectlanguage [0]{\@gobble}%
\providecommand \bibinfo  [0]{\@secondoftwo}%
\providecommand \bibfield  [0]{\@secondoftwo}%
\providecommand \translation [1]{[#1]}%
\providecommand \BibitemOpen [0]{}%
\providecommand \bibitemStop [0]{}%
\providecommand \bibitemNoStop [0]{.\EOS\space}%
\providecommand \EOS [0]{\spacefactor3000\relax}%
\providecommand \BibitemShut  [1]{\csname bibitem#1\endcsname}%
\let\auto@bib@innerbib\@empty
%</preamble>
\bibitem [{\citenamefont {Leggett}\ \emph {et~al.}(2006)\citenamefont {Leggett}
  \emph {et~al.}}]{Leggett2006quantum}%
  \BibitemOpen
  \bibfield  {author} {\bibinfo {author} {\bibfnamefont {A.~J.}\ \bibnamefont
  {Leggett}},\ }\href@noop {} {\emph {\bibinfo {title} {Quantum
  Liquids: Bose Condensation and Cooper pairing in Condensed-Matter Systems}}}\
  (\bibinfo  {publisher} {Oxford University Press},\ \bibinfo {year}
  {2006})\BibitemShut {NoStop}%
\bibitem [{\citenamefont {Anderson}\ \emph {et~al.}(1995)\citenamefont
  {Anderson}, \citenamefont {Ensher}, \citenamefont {Matthews}, \citenamefont
  {Wieman},\ and\ \citenamefont {Cornell}}]{Anderson1995}%
  \BibitemOpen
  \bibfield  {author} {\bibinfo {author} {\bibfnamefont {M.~H.}\ \bibnamefont
  {Anderson}}, \bibinfo {author} {\bibfnamefont {J.~R.}\ \bibnamefont
  {Ensher}}, \bibinfo {author} {\bibfnamefont {M.~R.}\ \bibnamefont
  {Matthews}}, \bibinfo {author} {\bibfnamefont {C.~E.}\ \bibnamefont
  {Wieman}},\ and\ \bibinfo {author} {\bibfnamefont {E.~A.}\ \bibnamefont
  {Cornell}},\ }\bibfield  {title} {\bibinfo {title} {{O}bservation of
  {B}ose-{E}instein condensation in a dilute atomic vapor},\ }\href
  {https://doi.org/10.1126/science.269.5221.198} {\bibfield  {journal}
  {\bibinfo  {journal} {Science}\ }\textbf {\bibinfo {volume} {269}},\ \bibinfo
  {pages} {198} (\bibinfo {year} {1995})}\BibitemShut {NoStop}%
\bibitem [{\citenamefont {Davis}\ \emph {et~al.}(1995)\citenamefont {Davis},
  \citenamefont {Mewes}, \citenamefont {Andrews}, \citenamefont {van Druten},
  \citenamefont {Durfee}, \citenamefont {Kurn},\ and\ \citenamefont
  {Ketterle}}]{Davis1995}%
  \BibitemOpen
  \bibfield  {author} {\bibinfo {author} {\bibfnamefont {K.~B.}\ \bibnamefont
  {Davis}}, \bibinfo {author} {\bibfnamefont {M.~O.}\ \bibnamefont {Mewes}},
  \bibinfo {author} {\bibfnamefont {M.~R.}\ \bibnamefont {Andrews}}, \bibinfo
  {author} {\bibfnamefont {N.~J.}\ \bibnamefont {van Druten}}, \bibinfo
  {author} {\bibfnamefont {D.~S.}\ \bibnamefont {Durfee}}, \bibinfo {author}
  {\bibfnamefont {D.~M.}\ \bibnamefont {Kurn}},\ and\ \bibinfo {author}
  {\bibfnamefont {W.}~\bibnamefont {Ketterle}},\ }\bibfield  {title} {\bibinfo
  {title} {{B}ose-{E}instein {C}ondensation in a {G}as of {S}odium {A}toms},\
  }\href {https://doi.org/10.1103/PhysRevLett.75.3969} {\bibfield  {journal}
  {\bibinfo  {journal} {Phys. Rev. Lett.}\ }\textbf {\bibinfo {volume} {75}},\
  \bibinfo {pages} {3969} (\bibinfo {year} {1995})}\BibitemShut {NoStop}%
\bibitem [{\citenamefont {Keldysh}\ and\ \citenamefont
  {Kozlov}(1968)}]{Keldysh1968}%
  \BibitemOpen
  \bibfield  {author} {\bibinfo {author} {\bibfnamefont {L.~V.}\ \bibnamefont
  {Keldysh}}\ and\ \bibinfo {author} {\bibfnamefont {A.~N.}\ \bibnamefont
  {Kozlov}},\ }\bibfield  {title} {\bibinfo {title} {Collective properties of
  excitons in semiconductors},\ }\href
  {http://jetp.ac.ru/cgi-bin/dn/e_027_03_0521.pdf} {\bibfield  {journal}
  {\bibinfo  {journal} {Sov. Phys. JETP}\ }\textbf {\bibinfo {volume} {27}},\
  \bibinfo {pages} {521} (\bibinfo {year} {1968})}\BibitemShut {NoStop}%
\bibitem [{\citenamefont {Kasprzak}\ \emph {et~al.}(2006)\citenamefont
  {Kasprzak}, \citenamefont {Richard}, \citenamefont {Kundermann},
  \citenamefont {Baas}, \citenamefont {Jeambrun}, \citenamefont {Keeling},
  \citenamefont {Marchetti}, \citenamefont {Szyma{\'{n}}ska}, \citenamefont
  {Andr{\'e}}, \citenamefont {Staehli}, \citenamefont {Savona}, \citenamefont
  {Littlewood}, \citenamefont {Deveaud},\ and\ \citenamefont
  {Dang}}]{Kasprzak2006}%
  \BibitemOpen
  \bibfield  {author} {\bibinfo {author} {\bibfnamefont {J.}~\bibnamefont
  {Kasprzak}}, \bibinfo {author} {\bibfnamefont {M.}~\bibnamefont {Richard}},
  \bibinfo {author} {\bibfnamefont {S.}~\bibnamefont {Kundermann}}, \bibinfo
  {author} {\bibfnamefont {A.}~\bibnamefont {Baas}}, \bibinfo {author}
  {\bibfnamefont {P.}~\bibnamefont {Jeambrun}}, \bibinfo {author}
  {\bibfnamefont {J.~M.~J.}\ \bibnamefont {Keeling}}, \bibinfo {author}
  {\bibfnamefont {F.~M.}\ \bibnamefont {Marchetti}}, \bibinfo {author}
  {\bibfnamefont {M.~H.}\ \bibnamefont {Szyma{\'{n}}ska}}, \bibinfo {author}
  {\bibfnamefont {R.}~\bibnamefont {Andr{\'e}}}, \bibinfo {author}
  {\bibfnamefont {J.~L.}\ \bibnamefont {Staehli}}, \bibinfo {author}
  {\bibfnamefont {V.}~\bibnamefont {Savona}}, \bibinfo {author} {\bibfnamefont
  {P.~B.}\ \bibnamefont {Littlewood}}, \bibinfo {author} {\bibfnamefont
  {B.}~\bibnamefont {Deveaud}},\ and\ \bibinfo {author} {\bibfnamefont {L.~S.}\
  \bibnamefont {Dang}},\ }\bibfield  {title} {\bibinfo {title}
  {Bose--{E}instein condensation of exciton polaritons},\ }\href
  {https://doi.org/10.1038/nature05131} {\bibfield  {journal} {\bibinfo
  {journal} {Nature (London)}\ }\textbf {\bibinfo {volume} {443}},\ \bibinfo
  {pages} {409} (\bibinfo {year} {2006})}\BibitemShut {NoStop}%
\bibitem [{\citenamefont {Demokritov}\ \emph {et~al.}(2006)\citenamefont
  {Demokritov}, \citenamefont {Demidov}, \citenamefont {Dzyapko}, \citenamefont
  {Melkov}, \citenamefont {Serga}, \citenamefont {Hillebrands},\ and\
  \citenamefont {Slavin}}]{Demokritov2006}%
  \BibitemOpen
  \bibfield  {author} {\bibinfo {author} {\bibfnamefont {S.~O.}\ \bibnamefont
  {Demokritov}}, \bibinfo {author} {\bibfnamefont {V.~E.}\ \bibnamefont
  {Demidov}}, \bibinfo {author} {\bibfnamefont {O.}~\bibnamefont {Dzyapko}},
  \bibinfo {author} {\bibfnamefont {G.~A.}\ \bibnamefont {Melkov}}, \bibinfo
  {author} {\bibfnamefont {A.~A.}\ \bibnamefont {Serga}}, \bibinfo {author}
  {\bibfnamefont {B.}~\bibnamefont {Hillebrands}},\ and\ \bibinfo {author}
  {\bibfnamefont {A.~N.}\ \bibnamefont {Slavin}},\ }\bibfield  {title}
  {\bibinfo {title} {{B}ose--{E}instein condensation of quasi-equilibrium
  magnons at room temperature under pumping},\ }\href
  {https://doi.org/10.1038/nature05117} {\bibfield  {journal} {\bibinfo
  {journal} {Nature (London)}\ }\textbf {\bibinfo {volume} {443}},\ \bibinfo
  {pages} {430} (\bibinfo {year} {2006})}\BibitemShut {NoStop}%
\bibitem [{\citenamefont {Proukakis}\ \emph {et~al.}(2017)\citenamefont
  {Proukakis}, \citenamefont {Snoke},\ and\ \citenamefont
  {Littlewood}}]{Proukakis2017BEC}%
  \BibitemOpen
  \bibinfo {editor} \href
  {https://doi.org/10.1017/9781316084366} {\emph {\bibinfo {title} {Universal
  {T}hemes of {B}ose-{E}instein {C}ondensation}}},\ edited by\ {\bibfnamefont {N.~P.}\ \bibnamefont {Proukakis}}, \bibinfo
  {editor} {\bibfnamefont {D.~W.}\ \bibnamefont {Snoke}},\ and\ \bibinfo
  {editor} {\bibfnamefont {P.~B.}\ \bibnamefont {Littlewood}} (\bibinfo  {publisher}
  {Cambridge University Press},\ \bibinfo {address} {Cambridge},\ \bibinfo
  {year} {2017})\BibitemShut {NoStop}%
\bibitem [{\citenamefont {Fisher}\ \emph {et~al.}(1989)\citenamefont {Fisher},
  \citenamefont {Weichman}, \citenamefont {Grinstein},\ and\ \citenamefont
  {Fisher}}]{Fisher1989}%
  \BibitemOpen
  \bibfield  {author} {\bibinfo {author} {\bibfnamefont {M.~P.~A.}\
  \bibnamefont {Fisher}}, \bibinfo {author} {\bibfnamefont {P.~B.}\
  \bibnamefont {Weichman}}, \bibinfo {author} {\bibfnamefont {G.}~\bibnamefont
  {Grinstein}},\ and\ \bibinfo {author} {\bibfnamefont {D.~S.}\ \bibnamefont
  {Fisher}},\ }\bibfield  {title} {\bibinfo {title} {Boson localization and the
  superfluid-insulator transition},\ }\href
  {https://doi.org/10.1103/PhysRevB.40.546} {\bibfield  {journal} {\bibinfo
  {journal} {Phys. Rev. B}\ }\textbf {\bibinfo {volume} {40}},\ \bibinfo
  {pages} {546} (\bibinfo {year} {1989})}\BibitemShut {NoStop}%
\bibitem [{\citenamefont {Greiner}\ \emph {et~al.}(2002)\citenamefont
  {Greiner}, \citenamefont {Mandel}, \citenamefont {Esslinger}, \citenamefont
  {H{\"a}nsch},\ and\ \citenamefont {Bloch}}]{Greiner2002}%
  \BibitemOpen
  \bibfield  {author} {\bibinfo {author} {\bibfnamefont {M.}~\bibnamefont
  {Greiner}}, \bibinfo {author} {\bibfnamefont {O.}~\bibnamefont {Mandel}},
  \bibinfo {author} {\bibfnamefont {T.}~\bibnamefont {Esslinger}}, \bibinfo
  {author} {\bibfnamefont {T.~W.}\ \bibnamefont {H{\"a}nsch}},\ and\ \bibinfo
  {author} {\bibfnamefont {I.}~\bibnamefont {Bloch}},\ }\bibfield  {title}
  {\bibinfo {title} {Quantum phase transition from a superfluid to a {M}ott
  insulator in a gas of ultracold atoms},\ }\href
  {https://doi.org/10.1038/415039a} {\bibfield  {journal} {\bibinfo  {journal}
  {Nature (London)}\ }\textbf {\bibinfo {volume} {415}},\ \bibinfo {pages} {39}
  (\bibinfo {year} {2002})}\BibitemShut {NoStop}%
\bibitem [{\citenamefont {Batyev}\ and\ \citenamefont
  {Braginski}(1984)}]{Batyev1984}%
  \BibitemOpen
  \bibfield  {author} {\bibinfo {author} {\bibfnamefont {E.~G.}\ \bibnamefont
  {Batyev}}\ and\ \bibinfo {author} {\bibfnamefont {L.~S.}\ \bibnamefont
  {Braginski}},\ }\bibfield  {title} {\bibinfo {title} {Antiferromagnet in a
  strong magnetic field: analogy with {B}ose gas},\ }\href
  {http://www.jetp.ac.ru/cgi-bin/dn/e_060_04_0781.pdf} {\bibfield  {journal}
  {\bibinfo  {journal} {Sov. Phys. JETP}\ }\textbf {\bibinfo {volume} {60}},\
  \bibinfo {pages} {781} (\bibinfo {year} {1984})}\BibitemShut {NoStop}%
\bibitem [{\citenamefont {Giamarchi}\ and\ \citenamefont
  {Tsvelik}(1999)}]{Giamarchi1999}%
  \BibitemOpen
  \bibfield  {author} {\bibinfo {author} {\bibfnamefont {T.}~\bibnamefont
  {Giamarchi}}\ and\ \bibinfo {author} {\bibfnamefont {A.~M.}\ \bibnamefont
  {Tsvelik}},\ }\bibfield  {title} {\bibinfo {title} {Coupled ladders in a
  magnetic field},\ }\href {https://doi.org/10.1103/PhysRevB.59.11398}
  {\bibfield  {journal} {\bibinfo  {journal} {Phys. Rev. B}\ }\textbf {\bibinfo
  {volume} {59}},\ \bibinfo {pages} {11398} (\bibinfo {year}
  {1999})}\BibitemShut {NoStop}%
\bibitem [{\citenamefont {Nikuni}\ \emph {et~al.}(2000)\citenamefont {Nikuni},
  \citenamefont {Oshikawa}, \citenamefont {Oosawa},\ and\ \citenamefont
  {Tanaka}}]{Nikuni2000}%
  \BibitemOpen
  \bibfield  {author} {\bibinfo {author} {\bibfnamefont {T.}~\bibnamefont
  {Nikuni}}, \bibinfo {author} {\bibfnamefont {M.}~\bibnamefont {Oshikawa}},
  \bibinfo {author} {\bibfnamefont {A.}~\bibnamefont {Oosawa}},\ and\ \bibinfo
  {author} {\bibfnamefont {H.}~\bibnamefont {Tanaka}},\ }\bibfield  {title}
  {\bibinfo {title} {{B}ose-{E}instein {C}ondensation of {D}ilute {M}agnons in
  {T}l{C}u{C}l$_{3}$},\ }\href {https://doi.org/10.1103/PhysRevLett.84.5868}
  {\bibfield  {journal} {\bibinfo  {journal} {Phys. Rev. Lett.}\ }\textbf
  {\bibinfo {volume} {84}},\ \bibinfo {pages} {5868} (\bibinfo {year}
  {2000})}\BibitemShut {NoStop}%
\bibitem [{\citenamefont {Giamarchi}\ \emph {et~al.}(2008)\citenamefont
  {Giamarchi}, \citenamefont {R{\"u}egg},\ and\ \citenamefont
  {Tchernyshyov}}]{Giamarchi2008}%
  \BibitemOpen
  \bibfield  {author} {\bibinfo {author} {\bibfnamefont {T.}~\bibnamefont
  {Giamarchi}}, \bibinfo {author} {\bibfnamefont {C.}~\bibnamefont
  {R{\"u}egg}},\ and\ \bibinfo {author} {\bibfnamefont {O.}~\bibnamefont
  {Tchernyshyov}},\ }\bibfield  {title} {\bibinfo {title} {Bose--{E}instein
  condensation in magnetic insulators},\ }\href
  {https://doi.org/10.1038/nphys893} {\bibfield  {journal} {\bibinfo  {journal}
  {Nat. Phys.}\ }\textbf {\bibinfo {volume} {4}},\ \bibinfo {pages} {198}
  (\bibinfo {year} {2008})}\BibitemShut {NoStop}%
\bibitem [{\citenamefont {Zapf}\ \emph {et~al.}(2014)\citenamefont {Zapf},
  \citenamefont {Jaime},\ and\ \citenamefont {Batista}}]{Zapf2014}%
  \BibitemOpen
  \bibfield  {author} {\bibinfo {author} {\bibfnamefont {V.}~\bibnamefont
  {Zapf}}, \bibinfo {author} {\bibfnamefont {M.}~\bibnamefont {Jaime}},\ and\
  \bibinfo {author} {\bibfnamefont {C.~D.}\ \bibnamefont {Batista}},\
  }\bibfield  {title} {\bibinfo {title} {{B}ose-{E}instein condensation in
  quantum magnets},\ }\href {https://doi.org/10.1103/RevModPhys.86.563}
  {\bibfield  {journal} {\bibinfo  {journal} {Rev. Mod. Phys.}\ }\textbf
  {\bibinfo {volume} {86}},\ \bibinfo {pages} {563} (\bibinfo {year}
  {2014})}\BibitemShut {NoStop}%
\bibitem [{\citenamefont {Lovesey}(1984)}]{Lovesey1984}%
  \BibitemOpen
  \bibfield  {author} {\bibinfo {author} {\bibfnamefont {S.~W.}\ \bibnamefont
  {Lovesey}},\ }\href@noop {} {\emph {\bibinfo {title} {Theory of {N}eutron
  {S}cattering from {C}ondensed {M}atter}}}\ (\bibinfo  {publisher} {Oxford University Press, Oxford},\
  \bibinfo {year} {1984})\BibitemShut {NoStop}%
\bibitem [{\citenamefont {Tokura}\ \emph {et~al.}(2014)\citenamefont {Tokura},
  \citenamefont {Seki},\ and\ \citenamefont {Nagaosa}}]{Tokura2014}%
  \BibitemOpen
  \bibfield  {author} {\bibinfo {author} {\bibfnamefont {Y.}~\bibnamefont
  {Tokura}}, \bibinfo {author} {\bibfnamefont {S.}~\bibnamefont {Seki}},\ and\
  \bibinfo {author} {\bibfnamefont {N.}~\bibnamefont {Nagaosa}},\ }\bibfield
  {title} {\bibinfo {title} {Multiferroics of spin origin},\ }\href
  {https://doi.org/10.1088/0034-4885/77/7/076501} {\bibfield  {journal}
  {\bibinfo  {journal} {Rep. Prog. Phys.}\ }\textbf {\bibinfo {volume} {77}},\
  \bibinfo {pages} {076501} (\bibinfo {year} {2014})}\BibitemShut {NoStop}%
\bibitem [{\citenamefont {Kim}\ \emph {et~al.}(2014)\citenamefont {Kim},
  \citenamefont {Khim}, \citenamefont {Chun}, \citenamefont {Jo}, \citenamefont
  {Balicas}, \citenamefont {Yi}, \citenamefont {Cheong}, \citenamefont
  {Harrison}, \citenamefont {Batista}, \citenamefont {Hoon~Han},\ and\
  \citenamefont {Hoon~Kim}}]{Kim2014}%
  \BibitemOpen
  \bibfield  {author} {\bibinfo {author} {\bibfnamefont {J.~W.}\ \bibnamefont
  {Kim}}, \bibinfo {author} {\bibfnamefont {S.}~\bibnamefont {Khim}}, \bibinfo
  {author} {\bibfnamefont {S.~H.}\ \bibnamefont {Chun}}, \bibinfo {author}
  {\bibfnamefont {Y.}~\bibnamefont {Jo}}, \bibinfo {author} {\bibfnamefont
  {L.}~\bibnamefont {Balicas}}, \bibinfo {author} {\bibfnamefont {H.~T.}\
  \bibnamefont {Yi}}, \bibinfo {author} {\bibfnamefont {S.-W.}\ \bibnamefont
  {Cheong}}, \bibinfo {author} {\bibfnamefont {N.}~\bibnamefont {Harrison}},
  \bibinfo {author} {\bibfnamefont {C.~D.}\ \bibnamefont {Batista}}, \bibinfo
  {author} {\bibfnamefont {J.~H}~\bibnamefont {Han}},\ and\ \bibinfo
  {author} {\bibfnamefont {K.~H}~\bibnamefont {Kim}},\ }\bibfield  {title}
  {\bibinfo {title} {Manifestation of magnetic quantum fluctuations in the
  dielectric properties of a multiferroic},\ }\href
  {https://doi.org/10.1038/ncomms5419} {\bibfield  {journal} {\bibinfo
  {journal} {Nat. Commun.}\ }\textbf {\bibinfo {volume} {5}},\ \bibinfo {pages}
  {4419} (\bibinfo {year} {2014})}\BibitemShut {NoStop}%
\bibitem [{\citenamefont {Arima}(2007)}]{Arima2007}%
  \BibitemOpen
  \bibfield  {author} {\bibinfo {author} {\bibfnamefont {T.-h.}\ \bibnamefont
  {Arima}},\ }\bibfield  {title} {\bibinfo {title} {Ferroelectricity induced
  by proper-screw type magnetic order},\ }\href
  {https://doi.org/10.1143/JPSJ.76.073702} {\bibfield  {journal} {\bibinfo
  {journal} {J. Phys. Soc. Jpn.}\ }\textbf {\bibinfo {volume} {76}},\ \bibinfo
  {pages} {073702} (\bibinfo {year} {2007})}\BibitemShut {NoStop}%
\bibitem [{\citenamefont {Jia}\ \emph {et~al.}(2007)\citenamefont {Jia},
  \citenamefont {Onoda}, \citenamefont {Nagaosa},\ and\ \citenamefont
  {Han}}]{Jia2007}%
  \BibitemOpen
  \bibfield  {author} {\bibinfo {author} {\bibfnamefont {C.}~\bibnamefont
  {Jia}}, \bibinfo {author} {\bibfnamefont {S.}~\bibnamefont {Onoda}}, \bibinfo
  {author} {\bibfnamefont {N.}~\bibnamefont {Nagaosa}},\ and\ \bibinfo {author}
  {\bibfnamefont {J.~H.}\ \bibnamefont {Han}},\ }\bibfield  {title} {\bibinfo
  {title} {Microscopic theory of spin-polarization coupling in multiferroic
  transition metal oxides},\ }\href
  {https://doi.org/10.1103/PhysRevB.76.144424} {\bibfield  {journal} {\bibinfo
  {journal} {Phys. Rev. B}\ }\textbf {\bibinfo {volume} {76}},\ \bibinfo
  {pages} {144424} (\bibinfo {year} {2007})}\BibitemShut {NoStop}%
\bibitem [{\citenamefont {Schrettle}\ \emph {et~al.}(2013)\citenamefont
  {Schrettle}, \citenamefont {Krohns}, \citenamefont {Lunkenheimer},
  \citenamefont {Loidl}, \citenamefont {Wulf}, \citenamefont {Yankova},\ and\
  \citenamefont {Zheludev}}]{Schrettle2013}%
  \BibitemOpen
  \bibfield  {author} {\bibinfo {author} {\bibfnamefont {F.}~\bibnamefont
  {Schrettle}}, \bibinfo {author} {\bibfnamefont {S.}~\bibnamefont {Krohns}},
  \bibinfo {author} {\bibfnamefont {P.}~\bibnamefont {Lunkenheimer}}, \bibinfo
  {author} {\bibfnamefont {A.}~\bibnamefont {Loidl}}, \bibinfo {author}
  {\bibfnamefont {E.}~\bibnamefont {Wulf}}, \bibinfo {author} {\bibfnamefont
  {T.}~\bibnamefont {Yankova}},\ and\ \bibinfo {author} {\bibfnamefont
  {A.}~\bibnamefont {Zheludev}},\ }\bibfield  {title} {\bibinfo {title}
  {Magnetic-field induced multiferroicity in a quantum critical frustrated spin
  liquid},\ }\href {https://doi.org/10.1103/PhysRevB.87.121105} {\bibfield
  {journal} {\bibinfo  {journal} {Phys. Rev. B}\ }\textbf {\bibinfo {volume}
  {87}},\ \bibinfo {pages} {121105(R)} (\bibinfo {year} {2013})}\BibitemShut
  {NoStop}%
\bibitem [{\citenamefont {Povarov}\ \emph {et~al.}(2015)\citenamefont
  {Povarov}, \citenamefont {Reichert}, \citenamefont {Wulf},\ and\
  \citenamefont {Zheludev}}]{Povarov2015}%
  \BibitemOpen
  \bibfield  {author} {\bibinfo {author} {\bibfnamefont {K.~Y.}\ \bibnamefont
  {Povarov}}, \bibinfo {author} {\bibfnamefont {A.}~\bibnamefont {Reichert}},
  \bibinfo {author} {\bibfnamefont {E.}~\bibnamefont {Wulf}},\ and\ \bibinfo
  {author} {\bibfnamefont {A.}~\bibnamefont {Zheludev}},\ }\bibfield  {title}
  {\bibinfo {title} {Giant dielectric nonlinearities at a magnetic
  {B}ose-{E}instein condensation},\ }\href
  {https://doi.org/10.1103/PhysRevB.92.140410} {\bibfield  {journal} {\bibinfo
  {journal} {Phys. Rev. B}\ }\textbf {\bibinfo {volume} {92}},\ \bibinfo
  {pages} {140410(R)} (\bibinfo {year} {2015})}\BibitemShut {NoStop}%
\bibitem [{\citenamefont {Kimura}\ \emph {et~al.}(2016)\citenamefont {Kimura},
  \citenamefont {Kakihata}, \citenamefont {Sawada}, \citenamefont {Watanabe},
  \citenamefont {Matsumoto}, \citenamefont {Hagiwara},\ and\ \citenamefont
  {Tanaka}}]{Kimura2016}%
  \BibitemOpen
  \bibfield  {author} {\bibinfo {author} {\bibfnamefont {S.}~\bibnamefont
  {Kimura}}, \bibinfo {author} {\bibfnamefont {K.}~\bibnamefont {Kakihata}},
  \bibinfo {author} {\bibfnamefont {Y.}~\bibnamefont {Sawada}}, \bibinfo
  {author} {\bibfnamefont {K.}~\bibnamefont {Watanabe}}, \bibinfo {author}
  {\bibfnamefont {M.}~\bibnamefont {Matsumoto}}, \bibinfo {author}
  {\bibfnamefont {M.}~\bibnamefont {Hagiwara}},\ and\ \bibinfo {author}
  {\bibfnamefont {H.}~\bibnamefont {Tanaka}},\ }\bibfield  {title} {\bibinfo
  {title} {Ferroelectricity by {B}ose--{E}instein condensation in a quantum
  magnet},\ }\href {https://doi.org/10.1038/ncomms12822} {\bibfield  {journal}
  {\bibinfo  {journal} {Nat. Commun.}\ }\textbf {\bibinfo {volume} {7}},\
  \bibinfo {pages} {12822} (\bibinfo {year} {2016})}\BibitemShut {NoStop}%
\bibitem [{\citenamefont {Kimura}\ \emph {et~al.}(2017)\citenamefont {Kimura},
  \citenamefont {Kakihata}, \citenamefont {Sawada}, \citenamefont {Watanabe},
  \citenamefont {Matsumoto}, \citenamefont {Hagiwara},\ and\ \citenamefont
  {Tanaka}}]{Kimura2017}%
  \BibitemOpen
  \bibfield  {author} {\bibinfo {author} {\bibfnamefont {S.}~\bibnamefont
  {Kimura}}, \bibinfo {author} {\bibfnamefont {K.}~\bibnamefont {Kakihata}},
  \bibinfo {author} {\bibfnamefont {Y.}~\bibnamefont {Sawada}}, \bibinfo
  {author} {\bibfnamefont {K.}~\bibnamefont {Watanabe}}, \bibinfo {author}
  {\bibfnamefont {M.}~\bibnamefont {Matsumoto}}, \bibinfo {author}
  {\bibfnamefont {M.}~\bibnamefont {Hagiwara}},\ and\ \bibinfo {author}
  {\bibfnamefont {H.}~\bibnamefont {Tanaka}},\ }\bibfield  {title} {\bibinfo
  {title} {Magnetoelectric effect in the quantum spin gap system
  {T}l{C}u{C}l$_{3}$},\ }\href {https://doi.org/10.1103/PhysRevB.95.184420}
  {\bibfield  {journal} {\bibinfo  {journal} {Phys. Rev. B}\ }\textbf {\bibinfo
  {volume} {95}},\ \bibinfo {pages} {184420} (\bibinfo {year}
  {2017})}\BibitemShut {NoStop}%
\bibitem [{\citenamefont {Mostovoy}(2006)}]{Mostovoy2006}%
  \BibitemOpen
  \bibfield  {author} {\bibinfo {author} {\bibfnamefont {M.}~\bibnamefont
  {Mostovoy}},\ }\bibfield  {title} {\bibinfo {title} {Ferroelectricity in
  {S}piral {M}agnets},\ }\href {https://doi.org/10.1103/PhysRevLett.96.067601}
  {\bibfield  {journal} {\bibinfo  {journal} {Phys. Rev. Lett.}\ }\textbf
  {\bibinfo {volume} {96}},\ \bibinfo {pages} {067601} (\bibinfo {year}
  {2006})}\BibitemShut {NoStop}%
\bibitem [{\citenamefont {Katsura}\ \emph {et~al.}(2005)\citenamefont
  {Katsura}, \citenamefont {Nagaosa},\ and\ \citenamefont
  {Balatsky}}]{Katsura2005}%
  \BibitemOpen
  \bibfield  {author} {\bibinfo {author} {\bibfnamefont {H.}~\bibnamefont
  {Katsura}}, \bibinfo {author} {\bibfnamefont {N.}~\bibnamefont {Nagaosa}},\
  and\ \bibinfo {author} {\bibfnamefont {A.~V.}\ \bibnamefont {Balatsky}},\
  }\bibfield  {title} {\bibinfo {title} {Spin {C}urrent and {M}agnetoelectric
  {E}ffect in {N}oncollinear {M}agnets},\ }\href
  {https://doi.org/10.1103/PhysRevLett.95.057205} {\bibfield  {journal}
  {\bibinfo  {journal} {Phys. Rev. Lett.}\ }\textbf {\bibinfo {volume} {95}},\
  \bibinfo {pages} {057205} (\bibinfo {year} {2005})}\BibitemShut {NoStop}%
\bibitem [{\citenamefont {Hase}\ \emph {et~al.}(2004)\citenamefont {Hase},
  \citenamefont {Kuroe}, \citenamefont {Ozawa}, \citenamefont {Suzuki},
  \citenamefont {Kitazawa}, \citenamefont {Kido},\ and\ \citenamefont
  {Sekine}}]{Hase2004}%
  \BibitemOpen
  \bibfield  {author} {\bibinfo {author} {\bibfnamefont {M.}~\bibnamefont
  {Hase}}, \bibinfo {author} {\bibfnamefont {H.}~\bibnamefont {Kuroe}},
  \bibinfo {author} {\bibfnamefont {K.}~\bibnamefont {Ozawa}}, \bibinfo
  {author} {\bibfnamefont {O.}~\bibnamefont {Suzuki}}, \bibinfo {author}
  {\bibfnamefont {H.}~\bibnamefont {Kitazawa}}, \bibinfo {author}
  {\bibfnamefont {G.}~\bibnamefont {Kido}},\ and\ \bibinfo {author}
  {\bibfnamefont {T.}~\bibnamefont {Sekine}},\ }\bibfield  {title} {\bibinfo
  {title} {Magnetic properties of {R}b$_{2}${C}u$_{2}${M}o$_{3}${O}$_{12}$
  including a one-dimensional spin-$1/2$ {H}eisenberg system with ferromagnetic
  first-nearest-neighbor and antiferromagnetic second-nearest-neighbor exchange
  interactions},\ }\href {https://doi.org/10.1103/PhysRevB.70.104426}
  {\bibfield  {journal} {\bibinfo  {journal} {Phys. Rev. B}\ }\textbf {\bibinfo
  {volume} {70}},\ \bibinfo {pages} {104426} (\bibinfo {year}
  {2004})}\BibitemShut {NoStop}%
\bibitem [{\citenamefont {Hase}\ \emph {et~al.}(2005)\citenamefont {Hase},
  \citenamefont {Ozawa}, \citenamefont {Suzuki}, \citenamefont {Kitazawa},
  \citenamefont {Kido}, \citenamefont {Kuroe},\ and\ \citenamefont
  {Sekine}}]{Hase2005}%
  \BibitemOpen
  \bibfield  {author} {\bibinfo {author} {\bibfnamefont {M.}~\bibnamefont
  {Hase}}, \bibinfo {author} {\bibfnamefont {K.}~\bibnamefont {Ozawa}},
  \bibinfo {author} {\bibfnamefont {O.}~\bibnamefont {Suzuki}}, \bibinfo
  {author} {\bibfnamefont {H.}~\bibnamefont {Kitazawa}}, \bibinfo {author}
  {\bibfnamefont {G.}~\bibnamefont {Kido}}, \bibinfo {author} {\bibfnamefont
  {H.}~\bibnamefont {Kuroe}},\ and\ \bibinfo {author} {\bibfnamefont
  {T.}~\bibnamefont {Sekine}},\ }\bibfield  {title} {\bibinfo {title}
  {Magnetism of ${A}_{2}${C}u$_{2}${M}o$_{3}${O}$_{12}$ (${A}$ = {R}b or {C}s):
  {M}odel compounds of a one-dimensional spin-1/2 {H}eisenberg system with
  ferromagnetic first-nearest-neighbor and antiferromagnetic
  second-nearest-neighbor interactions},\ }\href
  {https://doi.org/10.1063/1.1851675} {\bibfield  {journal} {\bibinfo
  {journal} {J. Appl. Phys. (Melville, NY)}\ }\textbf {\bibinfo {volume} {97}},\ \bibinfo
  {pages} {10B303} (\bibinfo {year} {2005})}\BibitemShut {NoStop}%
\bibitem [{\citenamefont {Yasui}\ \emph {et~al.}(2014)\citenamefont {Yasui},
  \citenamefont {Okazaki}, \citenamefont {Terasaki}, \citenamefont {Hase},
  \citenamefont {Hagihala}, \citenamefont {Masuda},\ and\ \citenamefont
  {Sakakibara}}]{Yasui2014}%
  \BibitemOpen
  \bibfield  {author} {\bibinfo {author} {\bibfnamefont {Y.}~\bibnamefont
  {Yasui}}, \bibinfo {author} {\bibfnamefont {R.}~\bibnamefont {Okazaki}},
  \bibinfo {author} {\bibfnamefont {I.}~\bibnamefont {Terasaki}}, \bibinfo
  {author} {\bibfnamefont {M.}~\bibnamefont {Hase}}, \bibinfo {author}
  {\bibfnamefont {M.}~\bibnamefont {Hagihala}}, \bibinfo {author}
  {\bibfnamefont {T.}~\bibnamefont {Masuda}},\ and\ \bibinfo {author}
  {\bibfnamefont {T.}~\bibnamefont {Sakakibara}},\ }\bibfield  {title}
  {\bibinfo {title} {Low temperature magnetic properties of frustrated
  quantum spin chain system {R}b$_{2}${C}u$_{2}${M}o$_{3}${O}$_{12}$},\
  }\href {https://doi.org/10.7566/JPSCP.3.014014} {\bibfield  {journal}
  {\bibinfo  {journal} {JPS Conf. Proc.}\ }\textbf {\bibinfo {volume} {3}},\
  \bibinfo {pages} {014014} (\bibinfo {year} {2014})}\BibitemShut {NoStop}%
\bibitem [{\citenamefont {Hayashida}\ \emph {et~al.}(2019)\citenamefont
  {Hayashida}, \citenamefont {Blosser}, \citenamefont {Povarov}, \citenamefont
  {Yan}, \citenamefont {Gvasaliya}, \citenamefont {Ponomaryov}, \citenamefont
  {Zvyagin},\ and\ \citenamefont {Zheludev}}]{Hayashida2019}%
  \BibitemOpen
  \bibfield  {author} {\bibinfo {author} {\bibfnamefont {S.}~\bibnamefont
  {Hayashida}}, \bibinfo {author} {\bibfnamefont {D.}~\bibnamefont {Blosser}},
  \bibinfo {author} {\bibfnamefont {K.~Y.}\ \bibnamefont {Povarov}}, \bibinfo
  {author} {\bibfnamefont {Z.}~\bibnamefont {Yan}}, \bibinfo {author}
  {\bibfnamefont {S.}~\bibnamefont {Gvasaliya}}, \bibinfo {author}
  {\bibfnamefont {A.~N.}\ \bibnamefont {Ponomaryov}}, \bibinfo {author}
  {\bibfnamefont {S.~A.}\ \bibnamefont {Zvyagin}},\ and\ \bibinfo {author}
  {\bibfnamefont {A.}~\bibnamefont {Zheludev}},\ }\bibfield  {title} {\bibinfo
  {title} {One- and three-dimensional quantum phase transitions and anisotropy
  in {R}b$_{2}${C}u$_{2}${M}o$_{3}${O}$_{12}$},\ }\href
  {https://doi.org/10.1103/PhysRevB.100.134427} {\bibfield  {journal} {\bibinfo
   {journal} {Phys. Rev. B}\ }\textbf {\bibinfo {volume} {100}},\ \bibinfo
  {pages} {134427} (\bibinfo {year} {2019})}\BibitemShut {NoStop}%
\bibitem [{\citenamefont {Yasui}\ \emph
  {et~al.}(2013{\natexlab{a}})\citenamefont {Yasui}, \citenamefont
  {Yanagisawa}, \citenamefont {Okazaki},\ and\ \citenamefont
  {Terasaki}}]{Yasui2013_1}%
  \BibitemOpen
  \bibfield  {author} {\bibinfo {author} {\bibfnamefont {Y.}~\bibnamefont
  {Yasui}}, \bibinfo {author} {\bibfnamefont {Y.}~\bibnamefont {Yanagisawa}},
  \bibinfo {author} {\bibfnamefont {R.}~\bibnamefont {Okazaki}},\ and\ \bibinfo
  {author} {\bibfnamefont {I.}~\bibnamefont {Terasaki}},\ }\bibfield  {title}
  {\bibinfo {title} {Dielectric anomaly in the quasi-one-dimensional frustrated
  spin-$\frac{1}{2}$ system
  {R}b$_{2}$({C}u$_{1-x}{{M}}_{x}$)$_{2}${M}o$_{3}${O}${}_{12}$ (${M}$ $=$ {N}i
  and {Z}n)},\ }\href {https://doi.org/10.1103/PhysRevB.87.054411} {\bibfield
  {journal} {\bibinfo  {journal} {Phys. Rev. B}\ }\textbf {\bibinfo {volume}
  {87}},\ \bibinfo {pages} {054411} (\bibinfo {year}
  {2013}{\natexlab{a}})}\BibitemShut {NoStop}%
\bibitem [{\citenamefont {Yasui}\ \emph
  {et~al.}(2013{\natexlab{b}})\citenamefont {Yasui}, \citenamefont
  {Yanagisawa}, \citenamefont {Okazaki}, \citenamefont {Terasaki},
  \citenamefont {Yamaguchi},\ and\ \citenamefont {Kimura}}]{Yasui2013_2}%
  \BibitemOpen
  \bibfield  {author} {\bibinfo {author} {\bibfnamefont {Y.}~\bibnamefont
  {Yasui}}, \bibinfo {author} {\bibfnamefont {Y.}~\bibnamefont {Yanagisawa}},
  \bibinfo {author} {\bibfnamefont {R.}~\bibnamefont {Okazaki}}, \bibinfo
  {author} {\bibfnamefont {I.}~\bibnamefont {Terasaki}}, \bibinfo {author}
  {\bibfnamefont {Y.}~\bibnamefont {Yamaguchi}},\ and\ \bibinfo {author}
  {\bibfnamefont {T.}~\bibnamefont {Kimura}},\ }\bibfield  {title} {\bibinfo
  {title} {Magnetic field induced ferroelectric transition of quasi
  one-dimensional frustrated quantum spin chain system
  {R}b$_{2}${C}u$_{2}${M}o$_{3}${O}$_{12}$},\ }\href
  {https://aip.scitation.org/doi/10.1063/1.4795844} {\bibfield  {journal}
  {\bibinfo  {journal} {J. Appl. Phys. (Melville, NY)}\ }\textbf {\bibinfo {volume} {113}},\
  \bibinfo {pages} {17D910} (\bibinfo {year} {2013}{\natexlab{b}})}\BibitemShut
  {NoStop}%
\bibitem [{\citenamefont {Reynolds}\ \emph {et~al.}(2019)\citenamefont
  {Reynolds}, \citenamefont {Mannig}, \citenamefont {Luetkens}, \citenamefont
  {Baines}, \citenamefont {Goko}, \citenamefont {Scheuermann}, \citenamefont
  {Keller}, \citenamefont {Bartkowiak}, \citenamefont {Fujimura}, \citenamefont
  {Yasui}, \citenamefont {Niedermayer},\ and\ \citenamefont
  {White}}]{Reynolds2019}%
  \BibitemOpen
  \bibfield  {author} {\bibinfo {author} {\bibfnamefont {N.}~\bibnamefont
  {Reynolds}}, \bibinfo {author} {\bibfnamefont {A.}~\bibnamefont {Mannig}},
  \bibinfo {author} {\bibfnamefont {H.}~\bibnamefont {Luetkens}}, \bibinfo
  {author} {\bibfnamefont {C.}~\bibnamefont {Baines}}, \bibinfo {author}
  {\bibfnamefont {T.}~\bibnamefont {Goko}}, \bibinfo {author} {\bibfnamefont
  {R.}~\bibnamefont {Scheuermann}}, \bibinfo {author} {\bibfnamefont
  {L.}~\bibnamefont {Keller}}, \bibinfo {author} {\bibfnamefont
  {M.}~\bibnamefont {Bartkowiak}}, \bibinfo {author} {\bibfnamefont
  {A.}~\bibnamefont {Fujimura}}, \bibinfo {author} {\bibfnamefont
  {Y.}~\bibnamefont {Yasui}}, \bibinfo {author} {\bibfnamefont
  {C.}~\bibnamefont {Niedermayer}},\ and\ \bibinfo {author} {\bibfnamefont
  {J.~S.}\ \bibnamefont {White}},\ }\bibfield  {title} {\bibinfo {title}
  {Magnetoelectric coupling without long-range magnetic order in the
  spin-$\frac{1}{2}$ multiferroic {R}b$_{2}${C}u$_{2}${M}o$_{3}${O}$_{12}$},\
  }\href {https://doi.org/10.1103/PhysRevB.99.214443} {\bibfield  {journal}
  {\bibinfo  {journal} {Phys. Rev. B}\ }\textbf {\bibinfo {volume} {99}},\
  \bibinfo {pages} {214443} (\bibinfo {year} {2019})}\BibitemShut {NoStop}%
\bibitem [{\citenamefont {Ueda}\ \emph {et~al.}(2020)\citenamefont {Ueda},
  \citenamefont {Onoda}, \citenamefont {Yamaguchi}, \citenamefont {Kimura},
  \citenamefont {Yoshizawa}, \citenamefont {Morioka}, \citenamefont {Hagiwara},
  \citenamefont {Hagihala}, \citenamefont {Soda}, \citenamefont {Masuda},
  \citenamefont {Sakakibara}, \citenamefont {Tomiyasu}, \citenamefont
  {Ohira-Kawamura}, \citenamefont {Nakajima}, \citenamefont {Kajimoto},
  \citenamefont {Nakamura}, \citenamefont {Inamura}, \citenamefont {Reynolds},
  \citenamefont {Frontzek}, \citenamefont {White}, \citenamefont {Hase},\ and\
  \citenamefont {Yasui}}]{Ueda2020}%
  \BibitemOpen
  \bibfield  {author} {\bibinfo {author} {\bibfnamefont {H.}~\bibnamefont
  {Ueda}}, \bibinfo {author} {\bibfnamefont {S.}~\bibnamefont {Onoda}},
  \bibinfo {author} {\bibfnamefont {Y.}~\bibnamefont {Yamaguchi}}, \bibinfo
  {author} {\bibfnamefont {T.}~\bibnamefont {Kimura}}, \bibinfo {author}
  {\bibfnamefont {D.}~\bibnamefont {Yoshizawa}}, \bibinfo {author}
  {\bibfnamefont {T.}~\bibnamefont {Morioka}}, \bibinfo {author} {\bibfnamefont
  {M.}~\bibnamefont {Hagiwara}}, \bibinfo {author} {\bibfnamefont
  {M.}~\bibnamefont {Hagihala}}, \bibinfo {author} {\bibfnamefont
  {M.}~\bibnamefont {Soda}}, \bibinfo {author} {\bibfnamefont {T.}~\bibnamefont
  {Masuda}}, \bibinfo {author} {\bibfnamefont {T.}~\bibnamefont {Sakakibara}},
  \bibinfo {author} {\bibfnamefont {K.}~\bibnamefont {Tomiyasu}}, \bibinfo
  {author} {\bibfnamefont {S.}~\bibnamefont {Ohira-Kawamura}}, \bibinfo
  {author} {\bibfnamefont {K.}~\bibnamefont {Nakajima}}, \bibinfo {author}
  {\bibfnamefont {R.}~\bibnamefont {Kajimoto}}, \bibinfo {author}
  {\bibfnamefont {M.}~\bibnamefont {Nakamura}}, \bibinfo {author}
  {\bibfnamefont {Y.}~\bibnamefont {Inamura}}, \bibinfo {author} {\bibfnamefont
  {N.}~\bibnamefont {Reynolds}}, \bibinfo {author} {\bibfnamefont
  {M.}~\bibnamefont {Frontzek}}, \bibinfo {author} {\bibfnamefont {J.~S.}\
  \bibnamefont {White}} {\emph {et~al}}.,\
  }\bibfield  {title} {\bibinfo {title} {Emergent spin-1 {H}aldane gap and
  ferroelectricity in a frustrated spin-$\frac{1}{2}$ ladder},\ }\href
  {https://doi.org/10.1103/PhysRevB.101.140408} {\bibfield  {journal} {\bibinfo
   {journal} {Phys. Rev. B}\ }\textbf {\bibinfo {volume} {101}},\ \bibinfo
  {pages} {140408(R)} (\bibinfo {year} {2020})}\BibitemShut {NoStop}%
\bibitem [{\citenamefont {Furukawa}\ \emph {et~al.}(2010)\citenamefont
  {Furukawa}, \citenamefont {Sato},\ and\ \citenamefont
  {Onoda}}]{Furukawa2010}%
  \BibitemOpen
  \bibfield  {author} {\bibinfo {author} {\bibfnamefont {S.}~\bibnamefont
  {Furukawa}}, \bibinfo {author} {\bibfnamefont {M.}~\bibnamefont {Sato}},\
  and\ \bibinfo {author} {\bibfnamefont {S.}~\bibnamefont {Onoda}},\ }\bibfield
   {title} {\bibinfo {title} {Chiral {O}rder and {E}lectromagnetic {D}ynamics
  in {O}ne-{D}imensional {M}ultiferroic {C}uprates},\ }\href
  {https://doi.org/10.1103/PhysRevLett.105.257205} {\bibfield  {journal}
  {\bibinfo  {journal} {Phys. Rev. Lett.}\ }\textbf {\bibinfo {volume} {105}},\
  \bibinfo {pages} {257205} (\bibinfo {year} {2010})}\BibitemShut {NoStop}%
\bibitem [{\citenamefont {Solodovnikov}\ and\ \citenamefont
  {Solodovnikova}(1997)}]{Solodovnikov1997}%
  \BibitemOpen
  \bibfield  {author} {\bibinfo {author} {\bibfnamefont {S.~F.}\ \bibnamefont
  {Solodovnikov}}\ and\ \bibinfo {author} {\bibfnamefont {Z.~A.}\ \bibnamefont
  {Solodovnikova}},\ }\bibfield  {title} {\bibinfo {title} {New structure type
  in the morphotropic series of
  $\mathrm{A}_{2}^{+}\mathrm{M}_{2}^{2+}(\mathrm{MoO}_{4})_{3}$: crystal
  structure of $\mathrm{Rb}_{2}\mathrm{Cu}_{2}(\mathrm{MoO}_{4})_{3}$},\ }\href
  {https://doi.org/10.1007/BF02763890} {\bibfield  {journal} {\bibinfo
  {journal} {J. Struct. Chem.}\ }\textbf {\bibinfo {volume} {38}},\ \bibinfo
  {pages} {765} (\bibinfo {year} {1997})}\BibitemShut {NoStop}%
\bibitem [{\citenamefont {Kuroe}\ \emph {et~al.}(2006)\citenamefont {Kuroe},
  \citenamefont {Hamasaki}, \citenamefont {Sekine}, \citenamefont {Hase},
  \citenamefont {Naka},\ and\ \citenamefont {Maeshima}}]{Kuroe2006}%
  \BibitemOpen
  \bibfield  {author} {\bibinfo {author} {\bibfnamefont {H.}~\bibnamefont
  {Kuroe}}, \bibinfo {author} {\bibfnamefont {T.}~\bibnamefont {Hamasaki}},
  \bibinfo {author} {\bibfnamefont {T.}~\bibnamefont {Sekine}}, \bibinfo
  {author} {\bibfnamefont {M.}~\bibnamefont {Hase}}, \bibinfo {author}
  {\bibfnamefont {T.}~\bibnamefont {Naka}},\ and\ \bibinfo {author}
  {\bibfnamefont {N.}~\bibnamefont {Maeshima}},\ }\bibfield  {title} {\bibinfo
  {title} {Effects of hydrostatic pressure on
  {R}b$_{2}${C}u$_{2}${M}o$_{3}${O}$_{12}$: a one-dimensional system with
  competing ferromagnetic and antiferromagnetic interactions},\ }\href
  {https://aip.scitation.org/doi/10.1063/1.2355063} {\emph {Low Temperature Physics: 24th International Conference on Low Temperature Physics, LT24}}, AIP Conference Proceedings (American Institute of Physics, Melville, NY, 2006) Vol.~850, p.~1049 \BibitemShut {NoStop}%
\bibitem [{\citenamefont {Hamasaki}\ \emph {et~al.}(2007)\citenamefont
  {Hamasaki}, \citenamefont {Kuroe}, \citenamefont {Sekine}, \citenamefont
  {Naka}, \citenamefont {Hase}, \citenamefont {Maeshima}, \citenamefont
  {Saiga},\ and\ \citenamefont {Uwatoko}}]{Hamasaki2007}%
  \BibitemOpen
  \bibfield  {author} {\bibinfo {author} {\bibfnamefont {T.}~\bibnamefont
  {Hamasaki}}, \bibinfo {author} {\bibfnamefont {H.}~\bibnamefont {Kuroe}},
  \bibinfo {author} {\bibfnamefont {T.}~\bibnamefont {Sekine}}, \bibinfo
  {author} {\bibfnamefont {T.}~\bibnamefont {Naka}}, \bibinfo {author}
  {\bibfnamefont {M.}~\bibnamefont {Hase}}, \bibinfo {author} {\bibfnamefont
  {N.}~\bibnamefont {Maeshima}}, \bibinfo {author} {\bibfnamefont
  {Y.}~\bibnamefont {Saiga}},\ and\ \bibinfo {author} {\bibfnamefont
  {Y.}~\bibnamefont {Uwatoko}},\ }\bibfield  {title} {\bibinfo {title} {Effects
  of high pressure on {A}$_{2}${C}u$_{2}${M}o$_{3}${O}$_{12}$ ({A} = {R}b,
  {C}s): {A} one-dimensional system with competing ferromagnetic and
  antiferromagnetic interactions},\ }\href
  {https://doi.org/https://doi.org/10.1016/j.jmmm.2006.10.370} {\bibfield
  {journal} {\bibinfo  {journal} {J. Magn. Magn. Mater.}\ }\textbf {\bibinfo
  {volume} {310}},\ \bibinfo {pages} {e394} (\bibinfo {year}
  {2007})}\BibitemShut {NoStop}%
\bibitem [{\citenamefont {Yagi}\ \emph {et~al.}(2017)\citenamefont {Yagi},
  \citenamefont {Matsui}, \citenamefont {Goto}, \citenamefont {Hase},\ and\
  \citenamefont {Sasaki}}]{Yagi2017}%
  \BibitemOpen
  \bibfield  {author} {\bibinfo {author} {\bibfnamefont {A.}~\bibnamefont
  {Yagi}}, \bibinfo {author} {\bibfnamefont {K.}~\bibnamefont {Matsui}},
  \bibinfo {author} {\bibfnamefont {T.}~\bibnamefont {Goto}}, \bibinfo {author}
  {\bibfnamefont {M.}~\bibnamefont {Hase}},\ and\ \bibinfo {author}
  {\bibfnamefont {T.}~\bibnamefont {Sasaki}},\ }\bibfield  {title} {\bibinfo
  {title} {{NMR} study on the competing spin chain
  {R}b$_{2}${C}u$_{2}${M}o$_{3}${O}$_{12}$},\ }\href
  {https://doi.org/10.1088/1742-6596/828/1/012016} {\bibfield  {journal}
  {\bibinfo  {journal} {J. Phys.: Conf. Ser.}\ }\textbf {\bibinfo {volume}
  {828}},\ \bibinfo {pages} {012016} (\bibinfo {year} {2017})}\BibitemShut
  {NoStop}%
\bibitem [{\citenamefont {Matsui}\ \emph {et~al.}(2017)\citenamefont {Matsui},
  \citenamefont {Yagi}, \citenamefont {Hoshino}, \citenamefont {Atarashi},
  \citenamefont {Hase}, \citenamefont {Sasaki},\ and\ \citenamefont
  {Goto}}]{Matsui2017}%
  \BibitemOpen
  \bibfield  {author} {\bibinfo {author} {\bibfnamefont {K.}~\bibnamefont
  {Matsui}}, \bibinfo {author} {\bibfnamefont {A.}~\bibnamefont {Yagi}},
  \bibinfo {author} {\bibfnamefont {Y.}~\bibnamefont {Hoshino}}, \bibinfo
  {author} {\bibfnamefont {S.}~\bibnamefont {Atarashi}}, \bibinfo {author}
  {\bibfnamefont {M.}~\bibnamefont {Hase}}, \bibinfo {author} {\bibfnamefont
  {T.}~\bibnamefont {Sasaki}},\ and\ \bibinfo {author} {\bibfnamefont
  {T.}~\bibnamefont {Goto}},\ }\bibfield  {title} {\bibinfo {title} {{R}b-{NMR}
  study of the quasi-one-dimensional competing spin-chain compound
  {R}b$_{2}${C}u$_{2}${M}o$_{3}${O}$_{12}$},\ }\href
  {https://doi.org/10.1103/PhysRevB.96.220402} {\bibfield  {journal} {\bibinfo
  {journal} {Phys. Rev. B}\ }\textbf {\bibinfo {volume} {96}},\ \bibinfo
  {pages} {220402(R)} (\bibinfo {year} {2017})}\BibitemShut {NoStop}%
\bibitem [{\citenamefont {Tomiyasu}\ \emph {et~al.}(2009)\citenamefont
  {Tomiyasu}, \citenamefont {Fujita}, \citenamefont {Kolesnikov}, \citenamefont
  {Bewley}, \citenamefont {Bull},\ and\ \citenamefont
  {Bennington}}]{Tomiyasu2009}%
  \BibitemOpen
  \bibfield  {author} {\bibinfo {author} {\bibfnamefont {K.}~\bibnamefont
  {Tomiyasu}}, \bibinfo {author} {\bibfnamefont {M.}~\bibnamefont {Fujita}},
  \bibinfo {author} {\bibfnamefont {A.~I.}\ \bibnamefont {Kolesnikov}},
  \bibinfo {author} {\bibfnamefont {R.~I.}\ \bibnamefont {Bewley}}, \bibinfo
  {author} {\bibfnamefont {M.~J.}\ \bibnamefont {Bull}},\ and\ \bibinfo
  {author} {\bibfnamefont {S.~M.}\ \bibnamefont {Bennington}},\ }\bibfield
  {title} {\bibinfo {title} {Conversion method of powder inelastic scattering
  data for one-dimensional systems},\ }\href
  {https://aip.scitation.org/doi/10.1063/1.3089566} {\bibfield  {journal}
  {\bibinfo  {journal} {Appl. Phys. Lett.}\ }\textbf {\bibinfo {volume} {94}},\
  \bibinfo {pages} {092502} (\bibinfo {year} {2009})}\BibitemShut {NoStop}%
\bibitem [{\citenamefont {Ohira-Kawamura}\ \emph {et~al.}(2018)\citenamefont
  {Ohira-Kawamura}, \citenamefont {Tomiyasu}, \citenamefont {Koda},
  \citenamefont {Sari}, \citenamefont {Asih}, \citenamefont {Yoon},
  \citenamefont {Watanabe},\ and\ \citenamefont {Nakajima}}]{Kawamura2018}%
  \BibitemOpen
  \bibfield  {author} {\bibinfo {author} {\bibfnamefont {S.}~\bibnamefont
  {Ohira-Kawamura}}, \bibinfo {author} {\bibfnamefont {K.}~\bibnamefont
  {Tomiyasu}}, \bibinfo {author} {\bibfnamefont {A.}~\bibnamefont {Koda}},
  \bibinfo {author} {\bibfnamefont {D.~P.}\ \bibnamefont {Sari}}, \bibinfo
  {author} {\bibfnamefont {R.}~\bibnamefont {Asih}}, \bibinfo {author}
  {\bibfnamefont {S.}~\bibnamefont {Yoon}}, \bibinfo {author} {\bibfnamefont
  {I.}~\bibnamefont {Watanabe}},\ and\ \bibinfo {author} {\bibfnamefont
  {K.}~\bibnamefont {Nakajima}},\ }\bibfield  {title} {\bibinfo {title}
  {Magnetic properties of one-dimensional quantum spin system
  {R}b$_{2}${C}u$_{2}${M}o$_{3}${O}$_{12}$ studied by muon spin
  relaxation},\ }\href {https://doi.org/10.7566/JPSCP.21.011007} {\bibfield
  {journal} {\bibinfo  {journal} {JPS Conf. Proc.}\ }\textbf {\bibinfo {volume}
  {21}},\ \bibinfo {pages} {011007} (\bibinfo {year} {2018})}\BibitemShut
  {NoStop}%
\bibitem [{\citenamefont {Kuroe}\ \emph {et~al.}(2011)\citenamefont {Kuroe},
  \citenamefont {Hosaka}, \citenamefont {Hachiuma}, \citenamefont {Sekine},
  \citenamefont {Hase}, \citenamefont {Oka}, \citenamefont {Ito}, \citenamefont
  {Eisaki}, \citenamefont {Fujisawa}, \citenamefont {Okubo},\ and\
  \citenamefont {Ohta}}]{Kuroe2011}%
  \BibitemOpen
  \bibfield  {author} {\bibinfo {author} {\bibfnamefont {H.}~\bibnamefont
  {Kuroe}}, \bibinfo {author} {\bibfnamefont {T.}~\bibnamefont {Hosaka}},
  \bibinfo {author} {\bibfnamefont {S.}~\bibnamefont {Hachiuma}}, \bibinfo
  {author} {\bibfnamefont {T.}~\bibnamefont {Sekine}}, \bibinfo {author}
  {\bibfnamefont {M.}~\bibnamefont {Hase}}, \bibinfo {author} {\bibfnamefont
  {K.}~\bibnamefont {Oka}}, \bibinfo {author} {\bibfnamefont {T.}~\bibnamefont
  {Ito}}, \bibinfo {author} {\bibfnamefont {H.}~\bibnamefont {Eisaki}},
  \bibinfo {author} {\bibfnamefont {M.}~\bibnamefont {Fujisawa}}, \bibinfo
  {author} {\bibfnamefont {S.}~\bibnamefont {Okubo}},\ and\ \bibinfo {author}
  {\bibfnamefont {H.}~\bibnamefont {Ohta}},\ }\bibfield  {title} {\bibinfo
  {title} {Electric polarization induced by {N}{\'e}el order without
  magnetic superlattice: experimental study of {C}u$_3${M}o$_2${O}$_9$
  and numerical study of a small spin cluster},\ }\href
  {https://doi.org/10.1143/JPSJ.80.083705} {\bibfield  {journal} {\bibinfo
  {journal} {J. Phys. Soc. Jpn.}\ }\textbf {\bibinfo {volume} {80}},\ \bibinfo
  {pages} {083705} (\bibinfo {year} {2011})}\BibitemShut {NoStop}%
\bibitem [{\citenamefont {Hamasaki}\ \emph {et~al.}(2008)\citenamefont
  {Hamasaki}, \citenamefont {Ide}, \citenamefont {Kuroe}, \citenamefont
  {Sekine}, \citenamefont {Hase}, \citenamefont {Tsukada},\ and\ \citenamefont
  {Sakakibara}}]{Hamasaki2008}%
  \BibitemOpen
  \bibfield  {author} {\bibinfo {author} {\bibfnamefont {T.}~\bibnamefont
  {Hamasaki}}, \bibinfo {author} {\bibfnamefont {T.}~\bibnamefont {Ide}},
  \bibinfo {author} {\bibfnamefont {H.}~\bibnamefont {Kuroe}}, \bibinfo
  {author} {\bibfnamefont {T.}~\bibnamefont {Sekine}}, \bibinfo {author}
  {\bibfnamefont {M.}~\bibnamefont {Hase}}, \bibinfo {author} {\bibfnamefont
  {I.}~\bibnamefont {Tsukada}},\ and\ \bibinfo {author} {\bibfnamefont
  {T.}~\bibnamefont {Sakakibara}},\ }\bibfield  {title} {\bibinfo {title}
  {Successive phase transitions to antiferromagnetic and weak-ferromagnetic
  long-range order in the quasi-one-dimensional antiferromagnet
  {C}u$_{3}${M}o$_{2}${O}$_{9}$},\ }\href
  {https://doi.org/10.1103/PhysRevB.77.134419} {\bibfield  {journal} {\bibinfo
  {journal} {Phys. Rev. B}\ }\textbf {\bibinfo {volume} {77}},\ \bibinfo
  {pages} {134419} (\bibinfo {year} {2008})}\BibitemShut {NoStop}%
\bibitem [{Note1()}]{Note1}%
  \BibitemOpen
  \bibinfo {note} {In this measurement the actual sample temperature is
  estimated to lag about 0.1~K behind the read out temperature due to the high
  heating rate}\BibitemShut {NoStop}%
\bibitem [{\citenamefont {Hikihara}\ \emph {et~al.}(2008)\citenamefont
  {Hikihara}, \citenamefont {Kecke}, \citenamefont {Momoi},\ and\ \citenamefont
  {Furusaki}}]{Hikihara2008}%
  \BibitemOpen
  \bibfield  {author} {\bibinfo {author} {\bibfnamefont {T.}~\bibnamefont
  {Hikihara}}, \bibinfo {author} {\bibfnamefont {L.}~\bibnamefont {Kecke}},
  \bibinfo {author} {\bibfnamefont {T.}~\bibnamefont {Momoi}},\ and\ \bibinfo
  {author} {\bibfnamefont {A.}~\bibnamefont {Furusaki}},\ }\bibfield  {title}
  {\bibinfo {title} {Vector chiral and multipolar orders in the
  spin-$\frac{1}{2}$ frustrated ferromagnetic chain in magnetic field},\ }\href
  {https://doi.org/10.1103/PhysRevB.78.144404} {\bibfield  {journal} {\bibinfo
  {journal} {Phys. Rev. B}\ }\textbf {\bibinfo {volume} {78}},\ \bibinfo
  {pages} {144404} (\bibinfo {year} {2008})}\BibitemShut {NoStop}%
\bibitem [{\citenamefont {Sudan}\ \emph {et~al.}(2009)\citenamefont {Sudan},
  \citenamefont {L\"uscher},\ and\ \citenamefont {L\"auchli}}]{Sudan2009}%
  \BibitemOpen
  \bibfield  {author} {\bibinfo {author} {\bibfnamefont {J.}~\bibnamefont
  {Sudan}}, \bibinfo {author} {\bibfnamefont {A.}~\bibnamefont {L\"uscher}},\
  and\ \bibinfo {author} {\bibfnamefont {A.~M.}\ \bibnamefont {L\"auchli}},\
  }\bibfield  {title} {\bibinfo {title} {Emergent multipolar spin correlations
  in a fluctuating spiral: {T}he frustrated ferromagnetic spin-$\frac{1}{2}$
  {H}eisenberg chain in a magnetic field},\ }\href
  {https://doi.org/10.1103/PhysRevB.80.140402} {\bibfield  {journal} {\bibinfo
  {journal} {Phys. Rev. B}\ }\textbf {\bibinfo {volume} {80}},\ \bibinfo
  {pages} {140402(R)} (\bibinfo {year} {2009})}\BibitemShut {NoStop}%
\bibitem [{\citenamefont {Sato}\ \emph {et~al.}(2013)\citenamefont {Sato},
  \citenamefont {Hikihara},\ and\ \citenamefont {Momoi}}]{Sato2013}%
  \BibitemOpen
  \bibfield  {author} {\bibinfo {author} {\bibfnamefont {M.}~\bibnamefont
  {Sato}}, \bibinfo {author} {\bibfnamefont {T.}~\bibnamefont {Hikihara}},\
  and\ \bibinfo {author} {\bibfnamefont {T.}~\bibnamefont {Momoi}},\ }\bibfield
   {title} {\bibinfo {title} {Spin-{N}ematic and {S}pin-{D}ensity-{W}ave
  {O}rders in {S}patially {A}nisotropic {F}rustrated {M}agnets in a {M}agnetic
  {F}ield},\ }\href {https://doi.org/10.1103/PhysRevLett.110.077206} {\bibfield
   {journal} {\bibinfo  {journal} {Phys. Rev. Lett.}\ }\textbf {\bibinfo
  {volume} {110}},\ \bibinfo {pages} {077206} (\bibinfo {year}
  {2013})}\BibitemShut {NoStop}%
\bibitem [{\citenamefont {Ueda}\ and\ \citenamefont {Momoi}(2013)}]{Ueda2013}%
  \BibitemOpen
  \bibfield  {author} {\bibinfo {author} {\bibfnamefont {H.~T.}\ \bibnamefont
  {Ueda}}\ and\ \bibinfo {author} {\bibfnamefont {T.}~\bibnamefont {Momoi}},\
  }\bibfield  {title} {\bibinfo {title} {Nematic phase and phase separation
  near saturation field in frustrated ferromagnets},\ }\href
  {https://doi.org/10.1103/PhysRevB.87.144417} {\bibfield  {journal} {\bibinfo
  {journal} {Phys. Rev. B}\ }\textbf {\bibinfo {volume} {87}},\ \bibinfo
  {pages} {144417} (\bibinfo {year} {2013})}\BibitemShut {NoStop}%
\bibitem [{\citenamefont {Starykh}\ and\ \citenamefont
  {Balents}(2014)}]{Starykh2014}%
  \BibitemOpen
  \bibfield  {author} {\bibinfo {author} {\bibfnamefont {O.~A.}\ \bibnamefont
  {Starykh}}\ and\ \bibinfo {author} {\bibfnamefont {L.}~\bibnamefont
  {Balents}},\ }\bibfield  {title} {\bibinfo {title} {Excitations and
  quasi-one-dimensionality in field-induced nematic and spin density wave
  states},\ }\href {https://doi.org/10.1103/PhysRevB.89.104407} {\bibfield
  {journal} {\bibinfo  {journal} {Phys. Rev. B}\ }\textbf {\bibinfo {volume}
  {89}},\ \bibinfo {pages} {104407} (\bibinfo {year} {2014})}\BibitemShut
  {NoStop}%
\bibitem [{\citenamefont {Matsumoto}\ \emph {et~al.}(2002)\citenamefont
  {Matsumoto}, \citenamefont {Normand}, \citenamefont {Rice},\ and\
  \citenamefont {Sigrist}}]{Matsumoto2002}%
  \BibitemOpen
  \bibfield  {author} {\bibinfo {author} {\bibfnamefont {M.}~\bibnamefont
  {Matsumoto}}, \bibinfo {author} {\bibfnamefont {B.}~\bibnamefont {Normand}},
  \bibinfo {author} {\bibfnamefont {T.~M.}\ \bibnamefont {Rice}},\ and\
  \bibinfo {author} {\bibfnamefont {M.}~\bibnamefont {Sigrist}},\ }\bibfield
  {title} {\bibinfo {title} {Magnon {D}ispersion in the {F}ield-{I}nduced
  {M}agnetically {O}rdered {P}hase of {T}l{C}u{C}l$_3$},\ }\href
  {https://doi.org/10.1103/PhysRevLett.89.077203} {\bibfield  {journal}
  {\bibinfo  {journal} {Phys. Rev. Lett.}\ }\textbf {\bibinfo {volume} {89}},\
  \bibinfo {pages} {077203} (\bibinfo {year} {2002})}\BibitemShut {NoStop}%
\bibitem [{\citenamefont {H\"uvonen}\ \emph {et~al.}(2012)\citenamefont
  {H\"uvonen}, \citenamefont {Zhao}, \citenamefont {M{\aa}nsson}, \citenamefont
  {Yankova}, \citenamefont {Ressouche}, \citenamefont {Niedermayer},
  \citenamefont {Laver}, \citenamefont {Gvasaliya},\ and\ \citenamefont
  {Zheludev}}]{Huvonen2012}%
  \BibitemOpen
  \bibfield  {author} {\bibinfo {author} {\bibfnamefont {D.}~\bibnamefont
  {H\"uvonen}}, \bibinfo {author} {\bibfnamefont {S.}~\bibnamefont {Zhao}},
  \bibinfo {author} {\bibfnamefont {M.}~\bibnamefont {M{\aa}nsson}}, \bibinfo
  {author} {\bibfnamefont {T.}~\bibnamefont {Yankova}}, \bibinfo {author}
  {\bibfnamefont {E.}~\bibnamefont {Ressouche}}, \bibinfo {author}
  {\bibfnamefont {C.}~\bibnamefont {Niedermayer}}, \bibinfo {author}
  {\bibfnamefont {M.}~\bibnamefont {Laver}}, \bibinfo {author} {\bibfnamefont
  {S.~N.}\ \bibnamefont {Gvasaliya}},\ and\ \bibinfo {author} {\bibfnamefont
  {A.}~\bibnamefont {Zheludev}},\ }\bibfield  {title} {\bibinfo {title}
  {Field-induced criticality in a gapped quantum magnet with bond disorder},\
  }\href {https://doi.org/10.1103/PhysRevB.85.100410} {\bibfield  {journal}
  {\bibinfo  {journal} {Phys. Rev. B}\ }\textbf {\bibinfo {volume} {85}},\
  \bibinfo {pages} {100410(R)} (\bibinfo {year} {2012})}\BibitemShut {NoStop}%
\bibitem [{\citenamefont {Continentino}(2011)}]{Continentino2011}%
  \BibitemOpen
  \bibfield  {author} {\bibinfo {author} {\bibfnamefont {M.~A.}\ \bibnamefont
  {Continentino}},\ }\bibfield  {title} {\bibinfo {title} {Interplay of quantum
  and classical fluctuations near quantum critical points},\ }\href
  {https://doi.org/10.1007/s13538-011-0025-2} {\bibfield  {journal} {\bibinfo
  {journal} {Braz. J. Phys.}\ }\textbf {\bibinfo {volume} {41}},\ \bibinfo
  {pages} {201} (\bibinfo {year} {2011})}\BibitemShut {NoStop}%
\end{thebibliography}


\end{document}